\documentclass[a4paper, 11pt, draftclsnofoot, onecolumn]{IEEEtran}

\usepackage{epsfig}
\usepackage{color}
\usepackage{graphicx}
\usepackage{amssymb,latexsym,amsmath}
\usepackage[latin1]{inputenc}
\usepackage{enumerate}
\usepackage{url}
\usepackage{pstricks}
\usepackage{fullpage}
\usepackage{pstricks}
\usepackage{setspace}

\usepackage{amsfonts,epsfig,graphicx}
\usepackage{amsmath,amssymb,amsthm}

\usepackage{epsf}
\usepackage{fancyheadings}
\usepackage{graphics}
\usepackage{amsfonts}
\usepackage{amsmath}
\usepackage{psfrag}

\newcommand{\lp}{\left(}
\newcommand{\rp}{\right)}
\newcommand{\lb}{\left[}
\newcommand{\rb}{\right]}

\newcommand{\vnorm}[1]{\ensuremath{\|#1\|_2}}

\newcommand{\matsnorm}[2]{|\!|\!| #1 | \! | \!|_{{#2}}}
\newcommand{\diag}{\operatorname{diag}}

\newcommand{\var}[1]{\ensuremath{\operatorname{var}}\lp#1\rp}
\newcommand{\Exp}[1]{\ensuremath{\mathbb{E}\lb#1\rb}}
\newcommand{\Prob}[1]{\ensuremath{\mathbb{P}[#1]}}
\newcommand{\ind}[1]{\ensuremath{\mathbb{I}\left\{#1\right\}}}
\newcommand{\as}{\stackrel{\text{a.s.}}{\longrightarrow}}

\newcommand{\asn}{\ensuremath{n^\ast}}

\newcommand{\bth}{\ensuremath{\theta}}
\newcommand{\bath}{\ensuremath{\bar{\theta}}}
\newcommand{\lam}{\ensuremath{\lambda}}
\newcommand{\alp}{\ensuremath{\alpha}}
\newcommand{\eps}{\ensuremath{\epsilon}}

\newcommand{\etajl}{\ensuremath{\eta_j^{(l)}}}
\newcommand{\tetaijl}{\ensuremath{\widetilde{\eta}_{ij}^{(l)}}}
\newcommand{\vij}{\ensuremath{v_{ij}^{(l)}}}

\newcommand{\sig}{\ensuremath{\sigma^2}}

\newcommand{\tgam}{\ensuremath{\widetilde{\gamma}}}

\newcommand{\epsp}{\ensuremath{\epsilon^\prime}}

\newcommand{\bCp}{\ensuremath{{C}^\prime}}
\newcommand{\Cp}{\ensuremath{C^{\prime}}}

\newcommand{\diam}{\ensuremath{D_n}}

\newcommand{\The}[1]{\ensuremath{\Theta\lp#1\rp}}
\newcommand{\bO}[1]{\ensuremath{\mathcal{O}\lp#1\rp}}

\newcommand{\bone}{\vec{1}}
\newcommand{\bzero}{\vec{0}}

\newcommand{\bC}{\ensuremath{C}}

\newcommand{\bu}{\ensuremath{u}}

\newcommand{\bv}{\ensuremath{v}}
\newcommand{\bL}{\ensuremath{L}}
\newcommand{\bW}{\ensuremath{W}}
\newcommand{\bvp}{\ensuremath{v^\prime}}
\newcommand{\bbW}{\widebar{W}}

\newcommand{\bbL}{\widebar{L}}
\newcommand{\bLam}{\Lambda}
\newcommand{\bU}{\ensuremath{U}}
\newcommand{\tbU}{\widetilde{\ensuremath{U}}}
\newcommand{\tbLam}{\widetilde{\Lambda}}

\newcommand{\bI}{\ensuremath{I}}

\newcommand{\bbet}{{\ensuremath{\beta}}}

\newtheorem{theorem}{Theorem}
\newtheorem{lemma}[theorem]{Lemma}

\newcommand{\graph}{\ensuremath{\mathcal{G}}}
\newcommand{\vertex}{\ensuremath{\mathcal{V}}}
\newcommand{\edge}{\ensuremath{\mathcal{E}}}

\newcommand{\numnode}{\ensuremath{n}}

\newcommand{\truemean}{\ensuremath{\mu}}

\newcommand{\real}{\ensuremath{\mathbb{R}}}

\newlength{\widebarargwidth}
\newlength{\widebarargheight}
\newlength{\widebarargdepth}
\DeclareRobustCommand{\widebar}[1]{%
  \settowidth{\widebarargwidth}{\ensuremath{#1}}%
  \settoheight{\widebarargheight}{\ensuremath{#1}}%
  \settodepth{\widebarargdepth}{\ensuremath{#1}}%
  \addtolength{\widebarargwidth}{-0.3\widebarargheight}%
  \addtolength{\widebarargwidth}{-0.3\widebarargdepth}%
  \makebox[0pt][l]{\hspace{0.3\widebarargheight}%
    \hspace{0.3\widebarargdepth}%
    \addtolength{\widebarargheight}{0.3ex}%
    \rule[\widebarargheight]{0.95\widebarargwidth}{0.1ex}}%
  {#1}}

\newcommand{\thetacent}{\ensuremath{\widebar{\theta}}}

\newcommand{\defn}{\ensuremath{:  = }}

\newcommand{\Exs}{\ensuremath{\mathbb{E}}}

\newcommand{\MSE}{\ensuremath{\operatorname{MSE}}}

\newcommand{\qprob}{\ensuremath{\mathbb{Q}}}

\newcommand{\spread}{\ensuremath{\psi}}
\newcommand{\spreadinv}{\ensuremath{\spread^{-1}}}

\newcommand{\neigh}{\ensuremath{\mathcal{N}}}

\newcommand{\mytheta}[1]{\ensuremath{\theta(#1)}}
\newcommand{\Tfinal}{\ensuremath{T}}

\newcommand{\chanvar}{\ensuremath{\sigma^2}}

\newcommand{\newt}{\ensuremath{\tau}}
\newcommand{\contt}{\ensuremath{\zeta}}

\newcommand{\newtin}{\ensuremath{s}}
\newcommand{\InnerIt}{\ensuremath{M}}
\newcommand{\myinnerest}[1]{\ensuremath{\gamma(#1)}}
\newcommand{\Emyinnerest}[2]{\ensuremath{\gamma_{#1}(#2)}}

\long\def\comment#1{}

\newenvironment{carlist}
 {\begin{list}{$\bullet$}
 {\setlength{\topsep}{0in} \setlength{\partopsep}{0in}
  \setlength{\parsep}{0in} \setlength{\itemsep}{\parskip}
  \setlength{\leftmargin}{0.07in} \setlength{\rightmargin}{0.08in}
  \setlength{\listparindent}{0in} \setlength{\labelwidth}{0.08in}
  \setlength{\labelsep}{0.1in} \setlength{\itemindent}{0in}}}
 {\end{list}}

\newcommand{\bcar}{\begin{carlist}}
\newcommand{\ecar}{\end{carlist}}

\newcommand{\Tgraph}[1]{\ensuremath{\Tfinal_{\scriptsize{\operatorname{#1}}}}}

\newcommand{\bit}{\begin{itemize}}
\newcommand{\eit}{\end{itemize}}

\newcommand{\order}{\ensuremath{\mathcal{O}}}

\newcommand{\Path}{\ensuremath{\mathcal{P}}}

\newcommand{\inprod}[2]{\ensuremath{\langle #1 \, , \, #2 \rangle}}

\newcommand{\SpecLap}{\ensuremath{S}}
\newcommand{\SpecLapAve}{{\ensuremath{\widebar{\SpecLap}}}}
\newcommand{\SpecLapAves}{{\ensuremath{\bar{\SpecLap}}}}
\newcommand{\Lap}{\ensuremath{L}}
\newcommand{\LapAve}{{\ensuremath{\widebar{\Lap}}}}

\newcommand{\trace}{\ensuremath{\operatorname{trace}}}

\newcommand{\Term}{\ensuremath{F}}

\newcommand{\Ltil}{\ensuremath{\underline{L}}}

\newcommand{\cov}{\ensuremath{\operatorname{cov}}}

\newcommand{\TransMat}{\ensuremath{A}}
\newcommand{\entry}{\ensuremath{a}}
\newcommand{\StaDist}{\ensuremath{\pi}}
\newcommand{\FirNode}{\ensuremath{s}}
\newcommand{\SecNode}{\ensuremath{u}}
\newcommand{\ThrNode}{\ensuremath{w}}
\newcommand{\Mpath}{\ensuremath{\eta}}
\newcommand{\PoinCoef}{\ensuremath{\kappa}}
\newcommand{\NewEdges}{\ensuremath{E^\prime}}
\newcommand{\SquaCent}{\ensuremath{\mathcal{C}}}
\newcommand{\Enum}{\ensuremath{\mathcal{N}}}

\newcommand{\widgraph}[2]{\includegraphics[keepaspectratio,width=#1]{#2}}

\newcommand{\Nat}{\ensuremath{\mathbb{N}}}

\newcommand{\myparagraph}[1]{\noindent {\textbf{#1}}}

\newcommand{\SpecGap}{\ensuremath{\lam_2(\SpecLapAves)}}
\newcommand{\Ferr}{\ensuremath{e_1}}
\newcommand{\Serr}{\ensuremath{e_2}}
\newcommand{\invdelt}{\ensuremath{\frac{1}{\delta}}}
\newcommand{\VF}{\ensuremath{H}}
\newcommand{\Const}{\ensuremath{c_0}}
\newcommand{\ConsValue}{\ensuremath{\widetilde{\bth}}}

\newcommand{\MAX}[2]{\ensuremath{\max \left\{ #1 \, , \, #2\right\}}}
\newcommand{\deltp}{\ensuremath {\widetilde {\delta}}}

\makeatletter
\long\def\@makecaption#1#2{
        \vskip 0.8ex
        \setbox\@tempboxa\hbox{\small {\bf #1:} #2}
        \parindent 1.5em  
        \dimen0=\hsize
        \advance\dimen0 by -3em
        \ifdim \wd\@tempboxa >\dimen0
                \hbox to \hsize{
                        \parindent 0em
                        \hfil
                        \parbox{\dimen0}{\def\baselinestretch{0.96}\small
                                {\bf #1.} #2
                                }
                        \hfil}
        \else \hbox to \hsize{\hfil \box\@tempboxa \hfil}
        \fi
        }
\makeatother

\begin{document}

\title{\LARGE \bf \vspace*{0.0 in} Non-asymptotic analysis of an optimal algorithm for \\
network-constrained averaging with noisy links }

\author{
Nima Noorshams$^1$ and Martin J. Wainwright$^{1,2}$\\
Departments of Statistics$^2$ and\\
Electrical Engineering \& Computer Science$^1$, \\
University of California Berkeley, \\
\{nshams, wainwrig\}@eecs.berkeley.edu }

\maketitle
\setcounter{page}{1}

\begin{abstract}
The problem of network-constrained averaging is to compute the
average of a set of values distributed throughout a graph $G$ using
an algorithm that can pass messages only along graph edges. We study
this problem in the noisy setting, in which the communication along
each link is modeled by an additive white Gaussian noise channel. We
propose a two-phase decentralized algorithm, and we use stochastic
approximation methods in conjunction with the spectral graph theory
to provide concrete (non-asymptotic) bounds on the mean-squared
error. Having found such bounds, we analyze how the number of
iterations $T_G(\numnode; \delta)$ required to achieve mean-squared
error $\delta$ scales as a function of the graph topology and the
number of nodes $\numnode$. Previous work provided guarantees with
the number of iterations scaling inversely with the second smallest
eigenvalue of the Laplacian. This paper gives an algorithm that
reduces this graph dependence to the graph diameter, which is the
best scaling possible.
\end{abstract}

%


\section{Introduction}\label{sec:introduction}

The problem of network-constrained averaging is to compute the average
of a set of numbers distributed throughout a network, using an
algorithm that is allowed to pass messages only along edges of the
graph.  Motivating applications include sensor networks, in which
individual motes have limited memory and communication ability, and
massive databases and server farms, in which memory constraints
preclude storing all data at a central location.  In typical
applications, the average might represent a statistical estimate of
some physical quantity (e.g., temperature, pressure etc.), or an
intermediate quantity in a more complex algorithm (e.g., for
distributed optimization).  There is now an extensive literature on
network-averaging, consensus problems, as well as distributed
optimization and estimation (e.g., see the
papers~\cite{BoydEtal06,DimSarWai08,DeGroot74,Tsitsiklis84,KemEtal03,AyselEtal09,BenezitEtal10,
CatSayed10, LopSayed08, LopSayaed07}).  The bulk of the earlier work
has focused on the noiseless variant, in which communication between
nodes in the graph is assumed to be noiseless. A more recent line of
work has studied versions of the problem with noisy communication
links (e.g., see the
papers~\cite{Hatano05,FagZam07,RajWai08,AysalEtal08,HanGamal09,
KarMoura09, NazEtal09} and references therein).

The focus of this paper is a noisy version of network-constrained
averaging in which inter-node communication is modeled by an
additive white Gaussian noise (AWGN) channel. Given this randomness,
any algorithm is necessarily stochastic, and the corresponding
sequence of random variables can be analyzed in various ways.  The
simplest question to ask is whether the algorithm is
consistent---that is, does it compute an approximate average or
achieve consensus in an asymptotic sense for a given fixed graph?  A
more refined analysis seeks to provide information about this
convergence rate.  In this paper, we do so by posing the following
question: for a given algorithm, how does number of iterations
required to compute the average to within $\delta$-accuracy scale as
a function of the graph topology and number of nodes $\numnode$? For
obvious reasons, we refer to this as the \emph{network scaling} of
an algorithm, and we are interested in finding an algorithm that has
near-optimal scaling law.

The issue of network scaling has been studied by a number of authors
in the noiseless setting, in which the communication between nodes
is perfect.  Of particular relevance here is the work of Benezit et
al.~\cite{BenEtal08_con}, who in the case of perfect communication,
provided a scheme that has essentially optimal message scaling law
for random geometric graphs.  A portion of the method proposed in
this paper is inspired by their scheme, albeit with suitable
extensions to multiple paths that are essential in the noisy
setting.  The issue of network scaling has also been studied in the
noisy setting; in particular, past work by Rajagopal and
Wainwright~\cite{RajWai08} analyzed a damped version of the usual
consensus updates, and provided scalings of the iteration number as
a function of the graph topology and size. However, our new
algorithm has much better scaling than the method~\cite{RajWai08}.

The main contributions of this paper are the development of a novel
two-phase algorithm for network-constrained averaging with noise,
and establishing the near-optimality of its network scaling.  At a
high level, the outer phase of our algorithm produces a sequence of
iterates $\{\theta(\tau)\}_{\tau=0}^\infty$ based on a recursive
linear update with decaying step size, as in stochastic
approximation methods.  The system matrix in this update is a
time-varying and random quantity, whose structure is determined by
the updates within the inner phase. These inner rounds are based on
establishing multiple paths between pairs of nodes, and averaging
along them simultaneously.  By combining a careful analysis of the
spectral properties of this random matrix with stochastic
approximation theory, we prove that this two-phase algorithm
computes a $\delta$-accurate version of the average using a number
of iterations that grows with the graph diameter (up to logarithmic
factors).\footnote{The graph diameter is the minimal
  number of edges needed to connect any two pairs of nodes in the
  graph.}  As we discuss in more detail following the statement of our
main result, this result is optimal up to logarithmic factors, meaning
that no algorithm can be substantially better in terms of network
scaling.

The remainder of this paper is organized as follows.  We begin in
Section~\ref{sec:problem_formulation} with background and formulation
of the problem.  In Section~\ref{SecMain}, we describe our algorithm,
and state various theoretical guarantees on its performance.  We then
provide the proof of our main result in Section~\ref{SecProof}.
Section~\ref{SecSimResult} is devoted to some simulation results that
confirm the sharpness of our theoretical predictions. We conclude the
paper in Section~\ref{SecConclusion}. \\

\noindent {\bf{Notation:}} For the reader's convenience, we collect
here some notation used throughout the paper. The notation $f(n) =
\mathcal{O}(g(n))$ means that there exists some constant \mbox{$c
\in (0, \infty)$} and $n_0 \in \Nat$ such $f(n) \leq c g(n)$ for all
$n \geq n_0$, whereas $f(n) = \Omega(g(n))$ means that $f(n) \geq c'
g(n)$ for all $n \geq n_0$.  The notation $f(n) = \Theta(g(n))$
means that $f(n) = \mathcal{O}(g(n))$ and \mbox{$f(n) =
\Omega(g(n))$.} Given a symmetric matrix $A \in \real^{n \times n}$,
we denote its ordered sequence of eigenvalues by $\lam_1(A) \leq
\lam_2(A) \leq \ldots \leq \lam_n(A)$ and also its $l_2$-operator
norm by $\matsnorm{A}{2} = \sup_{\vnorm{v} = 1} \vnorm{Av}$. Finally
we use $\inprod{\cdot}{\cdot}$ to denote the Euclidean inner
product.


\section{Background and problem set-up}
\label{sec:problem_formulation}

\noindent We begin in this section by introducing necessary background
and setting up the problem more precisely.

\subsection{Network-constrained averaging}
\label{subsec:averaging_problem}

Consider a collection $\{\theta_i(0), \: i = 1, \ldots, \numnode \}$
of $\numnode$ numbers.  In statistical settings, these numbers would
be modeled as identically distributed (i.i.d.) draws from an unknown
distribution $\qprob$ with mean $\truemean$.  In a centralized
setting, a standard estimator for the mean is the sample average
\mbox{$\thetacent \, \defn \frac{1}{\numnode} \sum_{i=1}^\numnode
\theta_i(0)$.}  When all of the data can be aggregated at a central
location, then computation of $\thetacent$ is straightforward.  In
this paper, we consider the network-constrained version of this
estimation problem, modeled by an undirected graph $\graph =
(\vertex, \edge)$ that consists of a vertex set $\vertex = \{1,
\ldots, \numnode\}$, and a collection of edges $\edge$ joining pairs
of vertices.  For $i \in \vertex$, we view each measurement
$\theta_i(0)$ as associated with vertex $i$.  (For instance, in the
context of sensor networks, each vertex would contain a mote and
collect observations of the environment.)  The edge structure of the
graph enforces communication constraints on the processing: in
particular, the presence of edge $(i,j)$ indicates that it is
possible for sensors $i$ and $j$ to exchange information via a noisy
communication channel. Conversely, sensor pairs that are \emph{not}
joined by an edge are not permitted to communicate
directly.\footnote{Moreover, since the edges are undirected, there
is no difference between edge $(i,j)$ and $(j,i)$; moreover, we
exclude self-edges, meaning that $(i, i) \notin \edge$ for all $i
\in \vertex$.}  Every node has a synchronized internal clock, and
acts at discrete times $t = 1, 2, \cdots$.  For any given pair of
sensors $(i,j) \in \edge$, we assume that the message sent from $i$
to $j$ is perturbed by an independent identically distributed $N(0,
\chanvar)$ variate. Although this additive white Gaussian noise
(AWGN) model is more realistic than a noiseless model, it is
conceivable (as pointed out by one of the reviewers) that other
stochastic channel models might be more suitable for certain types
of sensor networks, and we leave this exploration for future
research.

Given this set-up, of interest to us are stochastic algorithms that
generate sequences $\{ \mytheta{t} \}_{t=0}^\infty$ of iterates
contained within $\real^\numnode$, and we require that the algorithm
be \emph{graph-respecting}, meaning that in each iteration, it is
allowed to send at most one message for each direction of every edge
$(i,j) \in \edge$.  At time $t$, we measure the distance between
$\mytheta{t}$ and the desired average $\thetacent$ via the average
(per node) mean-squared error, given by
\begin{align}
\label{EqnMSEDefinition} \MSE(\mytheta{t}) & \defn
\frac{1}{\numnode} \sum_{i=1}^\numnode \Exs[
  (\bth_i(t) - \thetacent)^2].
\end{align}

In this paper, our goal is for every node to compute the average
$\thetacent$ up to an error tolerance $\delta$. In addition, we
require almost sure consensus  among nodes, meaning
\begin{align*}
\Prob { \bth_ i (t) = \bth_ j (t) \quad \forall \; i, j = 1 , 2 ,
  \cdots, \numnode} \to 1 \quad \text{as}\ t \to \infty.
\end{align*}
Our primary goal is in characterizing the rate of convergence as a
function of the graph topology and the number of nodes, to which we
refer as the \emph{network-scaling function} of the algorithm. More
precisely, in order to study this network scaling, we consider
sequences of graphs $\{\graph_{\numnode} \}$ indexed by the number of
nodes $\numnode$.  For any given algorithm (defined for each graph
$\graph_\numnode$) and a fixed tolerance parameter $\delta > 0$, our
goal is to determine bounds on the quantity
\begin{align} \label{EqnDefStopTime}
\Tfinal_\graph(\numnode; \delta) & \defn \inf \big \{ t = 1, 2, \ldots
\, \mid \, \MSE(\mytheta{t}) \leq \delta \big \}.
\end{align}
Note that $\Tfinal_\graph(\numnode; \delta)$ is a stopping time, given
by the smallest number of iterations required to obtain mean-squared
error less than $\delta$ on a graph of type $\graph$ with $\numnode$
nodes.


\subsection{Graph topologies}

\begin{figure*}
\setlength{\unitlength}{.8mm} \centering
\begin{tabular}{cccccc}
\begin{picture}(50,50)
\put(10,10){\line(1,-1){10}}\put(20,0){\circle{1}}
\put(20,0){\line(1,0){14}}\put(34,0){\circle{1}}
\put(34,0){\line(1,1){10}}\put(44,10){\circle{1}}
\put(44,10){\line(0,1){14}}\put(44,24){\circle{1}}
\put(44,24){\line(-1,1){10}}\put(34,34){\circle{1}}
\put(34,34){\line(-1,0){14}}\put(20,34){\circle{1}}
\put(20,34){\line(-1,-1){10}}\put(10,24){\circle{1}}
\put(10,24){\line(0,-1){14}}\put(10,10){\circle{1}}
\end{picture} & \hspace*{.02in} &
\begin{picture}(50,50)
\put(0,0){\dashbox{1}(50,50)}
\multiput(10,0)(10,0){4}{\dashbox{1}(0,50)}
\multiput(0,10)(0,10){4}{\dashbox{1}(50,0)}
\multiput(0,0)(0,10){5}{\multiput(5,5)(10,0){5}{\circle*{1}}}
\multiput(5,5)(0,10){5}{\line(1,0){40}}
\multiput(5,5)(10,0){5}{\line(0,1){40}}
\end{picture} &  \hspace*{.02in} &
\setlength{\unitlength}{.8mm} \centering
\begin{picture}(50,50)
\put(0,0){\dashbox{1}(50,50)}
\multiput(10,0)(10,0){4}{\dashbox{1}(0,50)}
\multiput(0,10)(0,10){4}{\dashbox{1}(50,0)}
\multiput(0,0)(0,10){5}{\multiput(5,5)(10,0){5}{\circle*{1}}}
\put(17,13){\circle{1}}  \qbezier(17,33)(37,33)(37,13)
\put(17,13){\dashbox{1}(20,0)} \put(22,10){\mbox{$r(n)$}}
\put(17,13){\line(-1,3){2}}\put(15,19){\circle{1}}
\put(17,13){\line(3,1){12}}\put(29,17){\circle{1}}
\put(17,13){\line(1,1){5}}\put(22,18){\circle{1}}
\put(17,13){\line(1,3){5}}\put(22,28){\circle{1}}
\put(17,13){\line(0,-1){5}}\put(17,8){\circle{1}}
\put(17,13){\line(-1,-2){5}}\put(12,3){\circle{1}}
\put(17,13){\line(-1,0){10}}\put(7,13){\circle{1}}
\put(17,13){\line(-1,6){1.5}}\put(15.5,22){\circle{1}}
\put(17,13){\line(1,-2){5}}\put(22,3){\circle{1}}
\end{picture} & \\
(a) Single cycle.  & & (b) Two-dimensional grid.  & & (c)
Random geometric graph. \\
& & & &
\end{tabular}
\caption{Illustration of graph topologies.  (a) A single cycle graph.
  (b) Two-dimensional grid with four-nearest-neighbor connectivity.
  (c) Illustration of a random geometric graph (RGG). Two nodes are
  connected if their distance is less than $r(n)$. The solid circles
  represent the center of squares.  }
  \label{FigGraphs}
\end{figure*}
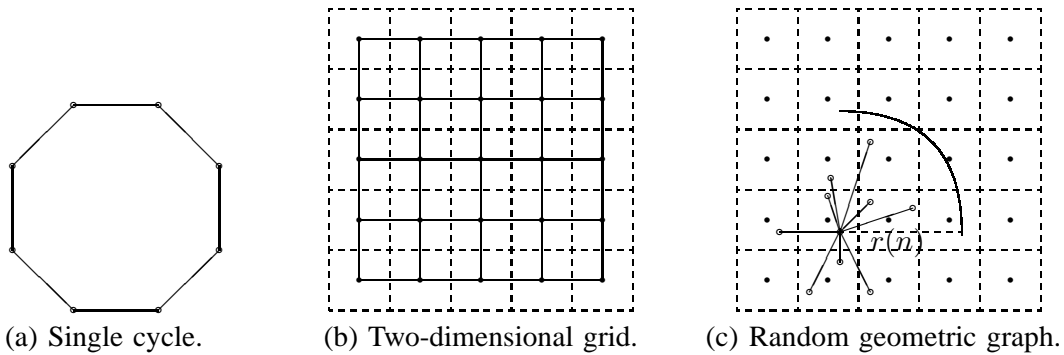

Of course, the question that we have posed will depend on the graph
type, and this paper analyzes three types of graphs, as shown in
Figure~\ref{FigGraphs}.  The first two graphs have regular topologies:
the single cycle graph in panel (a) is degree two-regular, and the
two-dimensional grid graph in panel (b) is degree four-regular.  In
addition, we also analyze an important class of random graphs with
irregular topology, namely the class of random geometric graphs.  As
illustrated in Figure~\ref{FigGraphs}(c), a random geometric graph
(RGG) in the plane is formed according by placing $\numnode$ nodes
uniformly at random in the unit square $[0,1] \times [0,1]$, and the
connecting two nodes if their Euclidean distance is less than some
radius $r(\numnode)$.  It is known that an RGG will be connected with
high probability as long as $r(\numnode) = \Omega(\sqrt{\frac{\log
    \numnode}{\numnode}})$; see Penrose~\cite{Pen03} for discussion of
this and other properties of random geometric graphs.

A key graph-theoretic parameter relevant to our analysis is the
\emph{graph diameter}, denoted by $\diam = \operatorname {diam}
(\graph _ \numnode)$. The path distance between any pair of nodes is
the length of the shortest path joining them in the graph, and by
definition, the graph diameter is the maximum path distance taken over
all node pairs in the graph.  It is straightforward to see that
\mbox{$\diam = \Theta(\numnode)$} for the single cycle graph, and that
\mbox{$\diam = \Theta(\sqrt{\numnode})$} for the two-dimensional grid.
For a random geometric graph with radius chosen to ensure
connectivity, it is known that $\diam=\The{\sqrt{\frac{\numnode
}{\log{\numnode}}}}$.

Finally, in order to simplify the routing problem explained later,
we divide the unit square into subregions (squares) of side length
$\sqrt{\frac{1}{n}}$ in case of grid, and for some constant $c > 0$,
of side length $\sqrt{c\frac{\log{n}}{n}}$ in case of RGG.  We
assume that each node knows its location and is aware of the center
of these $m^2$ subregions namely $({x}_i,{y}_j)$ $i,j=1,2,\cdots,m$,
where $m=\sqrt{n}$ for the regular grid, and
$m=\sqrt{\frac{n}{c\log{n}}}$ for the RGG. As a convention, we
assume that $(x_1,y_1)$ is the left bottom square, to which we refer
to as the first square. By construction, in a regular grid, each
square will contain one and only one node which is located at the
center of the square.  From known properties of RGGs~\cite{Pen03,
GupKum00}, each of the given subregions will contain at least one
node with high probability (w.h.p.).  Moreover, an RGG is regular
w.h.p, meaning that each square contains $\The{\log{n}}$ nodes (see
Lemma 1 in the paper~\cite{DimSarWai08}). Accordingly, in the
remainder of the paper, we assume without loss of generality that
any given RGG is regular. Note that by construction, the
transmission radius $r(n)$ is selected so that each node in each
square is connected to every other node in four adjacent squares.


\section{Algorithm and its properties}
\label{SecMain}

In this section we state our main result which is followed by a
detailed description of the proposed algorithm.

\subsection{Theoretical guarantees}

Our main result guarantees the existence of a graph-respecting
algorithm with desirable properties. Recall the definition of the
graph respecting scheme, as well as the definition of our AWGN
channel model given in Section~\ref{sec:problem_formulation}. In the
following statement, the quantity $\Const$ denotes a universal
constant, independent of $\numnode$, $\delta$, and $\sig$.
\begin{theorem}
\label{ThmMain}
For the communication model in which each link is an AWGN channel with
variance $\sigma^2$, there is a graph-respecting algorithm such that:
\begin{enumerate}[a)]
\item Nodes almost surely reach a consensus. More precisely, we have
\begin{align}\label{EqnConsensus}
\bth (t) \as \ConsValue \: \bone \quad \text{as $t \to \infty$},
\end{align}
for some $\ConsValue \in \real$.
\item After $\Tfinal = \Tgraph{\graph}(\numnode; \delta)$ iterations, the algorithm satisfy the following bounds on the $\MSE (\bth (\Tfinal))$:
\begin{enumerate} [i)]
\item For fixed tolerance $\delta > 0$ sufficiently small, we have $\MSE(\bth(\Tfinal)) \le 3 \: \sig
\delta$ after
\begin{align*}
\Tgraph{cyc}(\numnode; \delta) \, \le \, \Const \: \numnode \:
\MAX{\invdelt \log{\invdelt}} {\frac{\MSE (\bth(0))}{\sig \delta
^2}}
\end{align*}
iterations for a single cycle graph.
\item For fixed tolerance $\delta > 0$ sufficiently small, we have $\MSE(\bth(\Tfinal)) = \bO {\sig
\delta}$ after
\begin{align*}
\Tgraph{grid}(\numnode; \delta) \, \le \, \Const \: \sqrt{\numnode}
\: \MAX{\invdelt \log{\invdelt}} {\frac{\MSE (\bth(0))}{\sig \delta
^2}}
\end{align*}
  iterations for the regular grid in two dimensions.
\item Assume that $\delta = \frac {\deltp} {(\log{\numnode}) ^2}$, for some fixed $\deltp$ sufficiently small. Then we
have $\MSE(\bth(\Tfinal)) = \bO{ \sig \deltp}$ after
\begin{align*}
\Tgraph{RGG}(\numnode; \delta) \, \le \, \Const \: \sqrt {\numnode
(\log{\numnode}) ^3} \: \MAX { \frac {1} {\deltp} \log{\frac
{(\log{\numnode})^2} {\deltp}} } { \frac{\MSE (\bth(0))}{\sig \deltp
^2} }
\end{align*}
iterations for a regular random geometric graph.
\end{enumerate}
Here $\Const$ is some constant independent of $\numnode$, $\delta$,
and $\sig$, whose value may change from line to line.
\end{enumerate}

\end{theorem}
\myparagraph{Remarks:} A few comments are in order regarding the
interpretation of this result.  First, it is worth mentioning that
the quality of the different links does not have to be the same.
Similar arguments apply to the case where noises have different
variances. Second, although nodes almost surely reach a consensus,
as guaranteed in part (a), this consensus value is not necessarily
the same as the sample mean $\bath$. The choice of $\ConsValue$ is
intentional to emphasize this point. However, as guaranteed by part
(b), this consensus value is within $\sig \delta$ distance of the
actual sample mean. Since the sample mean itself represents a noisy
estimate of some underlying population quantity, there is little
point to computing it to arbitrary accuracy.  Third, it is
worthwhile comparing part (b) with previous results on network
scaling in the noisy setting. Rajagopal and
Wainwright~\cite{RajWai08} analyzed a simple set of damped updates,
and showed that $\Tgraph{cyc}(\numnode; \delta) = \bO{\numnode^2}$
for the single cycle, and that \mbox{$\Tgraph{grid}(\numnode) =
\bO{\numnode}$} for the two-dimensional grid.  By comparison, the
algorithm proposed here and our analysis thereof has removed factors
of $\numnode$ and $\sqrt{\numnode}$ from this scaling.

\subsection{Optimality of the results}

As we now discuss, the scalings in Theorem~\ref{ThmMain} are optimal
for the cases of cycle and grid and near-optimal (up to logarithmic
factor) for the case of RGG. In an adversarial setting, any
algorithm needs at least $\Omega(\diam)$ iterations, where $\diam$
denotes the graph diameter, in order to approximate the average;
otherwise, some node will fail to have any information from some
subset of other nodes (and their values can be set in a worst-case
manner). Theorem~\ref{ThmMain} provides upper bounds on the number
of iterations that, at most, are within logarithmic factors of the
diameter, and hence are also within logarithmic factors of the
optimal latency scaling law. For the graphs given here, the scalings
are also optimal in a non-adversarial setting, in which
$\{\theta_i(0)\}_{i=1}^\numnode$ are modeled as chosen i.i.d. from
some distribution.  Indeed, for a given node $j \in \vertex$, and
positive integer $t$, we let $\neigh(j;t)$ denote the depth $t$
neighborhood of $j$, meaning the set of nodes that are connected to
$j$ by a path of length at most $t$. We then define the graph
spreading function $\spread_{\graph}(t) = \min_{j \in \vertex}
|\neigh(j; t)|$.  Note that the function $\spread_\graph$ is
non-decreasing, so that we may define its inverse function
\mbox{$\spread_\graph^{-1}(s) = \inf \{ t \, \mid \spread_\graph(t)
\leq s \}$}. As some examples:
\begin{itemize}
\item for a cycle on $\numnode$ nodes, we have $\spread_\graph(t) = 2 t$,
and hence $\spreadinv_\graph(s) = s/2$.
\item for a $\numnode$-grid in two dimensions, we have the upper bound
  $\spread_\graph(t) \leq 2t^2$, and hence the lower bound
  $\spreadinv_\graph(s) \geq \sqrt{\frac{s}{2}}$.
\item for a random geometric graph (RGG), we have the upper bound
  $\spread_\graph(t)=\Theta(t^2\log{n})$, which implies the lower
  bound
  $\spreadinv_\graph(s)=\Theta\left(\sqrt{\frac{s}{\log{n}}}\right)$
\end{itemize}
After $t$ steps, a given node can gather the information of at most
$\spread_\graph(t)$ nodes.  For the average based on
$\spread_\graph(t)$ nodes to be comparable to $\thetacent$, we
require that $\spread_\graph(t)=\Omega(\numnode)$, and hence the
iteration number $t$ should be at least
$\Omega(\spreadinv_\graph(\numnode))$.  For the three graphs
considered here, this leads to the same conclusion, namely that
$\Omega(\diam)$ iterations are required. We note also that using
information-theoretic techniques, Ayaso et al.~\cite{AyasoEtal}
proved a lower bound on the number of iterations for a general graph
in terms of the Cheeger constant \cite{Chung97}. For the graphs
considered here, the Cheeger constant is of the order of the
diameter.


\subsection{Description of algorithm}

\begin{figure}[h]
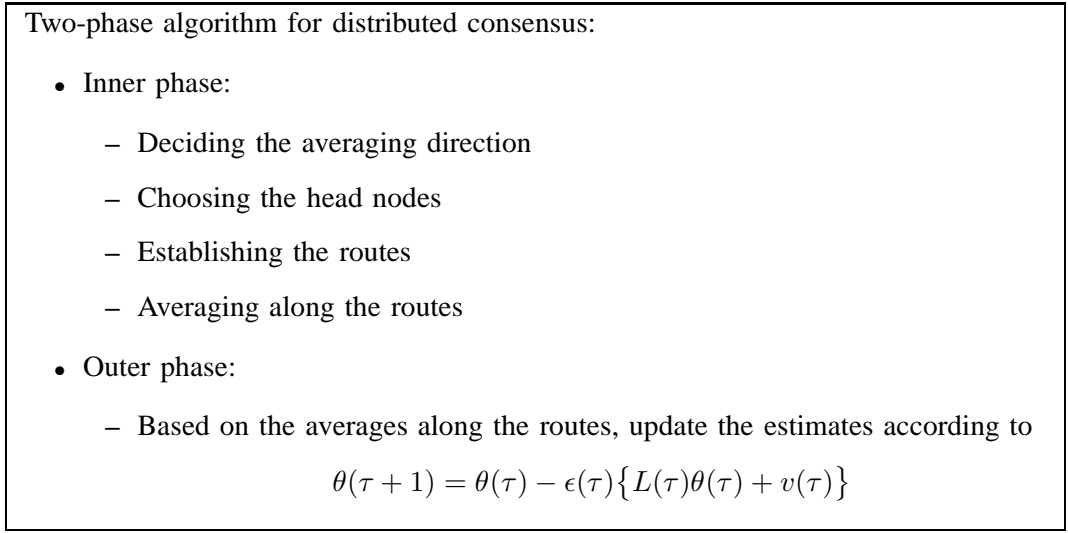

\begin{center}
\framebox[.82\textwidth]{\parbox{.79\textwidth}{ Two-phase algorithm
for distributed consensus:
\begin{itemize}
\vspace{.1in}
\item Inner phase:
\begin{itemize}
\vspace{.1in}
\item Deciding the averaging direction
\vspace{.1in}
\item Choosing the head nodes
\vspace{.1in}
\item Establishing the routes
\vspace{.1in}
\item Averaging along the routes
\end{itemize}
\vspace{.1in}
\item Outer phase:
\begin{itemize}
\vspace{.1in}
\item Based on the averages along the routes, update the estimates according to
\begin{align*}
\mytheta{\newt+1}=\mytheta{\newt}-\eps(\newt) \big \{ \bL(\newt)
\bth(\newt)+\bv(\newt) \big\}
\end{align*}
\end{itemize}
\end{itemize}
}} \caption{Basic operations of a two-phase algorithm for distributed
consensus.}
\label{Fig:BasicOperation}
\end{center}
\end{figure}

We now describe the algorithm that achieves the bounds stated in
Theorem~\ref{ThmMain}. At the highest level, the algorithm can be
divided into two types of phases: an inner phase, and an outer
phase. The outer phase produces a sequence of iterates
$\{\mytheta{\newt}\}$, where $\newt = 0, 1, 2, \ldots$ is the outer
time scale parameter. By design of the algorithm, each update of the
outer parameters requires a total of $\InnerIt$ message-passing rounds
(these rounds corresponding to the inner phase), where in each round
the algorithm can pass at most two messages per edge (one for each
direction). To put everything in a nutshell, the algorithm is based on
establishing multiple routes, averaging along them in an inner phase
and updating the estimates based on the noisy version of averages
along routes in an outer phase. Consequently, if we use the estimate
$\mytheta{\newt}$, then in the language of Theorem~\ref{ThmMain}, it
corresponds to $\Tfinal = \InnerIt \tau$ rounds of
message-passing. Our goal is to establish upper bounds on $\Tfinal$
that guarantee the MSE is $\order(\sig \delta)$.  Figure
\ref{Fig:BasicOperation} illustrates the basic operations of the
algorithm.

\subsubsection{Outer phase}
In the outer phase, we produce a sequence of iterates
$\{\mytheta{\newt}\}_{\newt=1}^\infty$ according to the recursive
update
\begin{equation}
\label{eq:outer_phase} \mytheta{\newt+1}=\mytheta{\newt}-\eps(\newt)
\big \{ \bL(\newt) \bth(\newt)+\bv(\newt) \big\}.
\end{equation}
Here $\{\eps(\newt) \}_{\newt = 1}^\infty$ is a sequence of positive
decreasing stepsizes. For a given precision, $\delta$, we set $\eps
(\newt) = 1 / (\frac{1}{\delta} + \newt)$. For each $\newt$, the
quantity $\bL(\newt) \in \real^{\numnode \times \numnode}$ is a
random matrix, whose structure is determined by the inner phase, and
$\bv(\newt) \in \real^\numnode$ is an additive Gaussian term, whose
structure is also determined in the inner phase.  As will become
clear in the sequel, even though $\bL$ and $\bv$ are dependent, they
are both independent of $\bth$. Moreover, given $\bL$, the random
vector $\bv$ is Gaussian with bounded variance.


\subsubsection{Inner phase}
\label{SubSubSecInnerPhase}

The inner phase is the core of the algorithm and it involves a
number of steps, as we describe here.  We use $\newtin = 1, 2,
\ldots, \InnerIt$ to index the iterations within any inner phase,
and use $\{\myinnerest{\newtin}\}_{\newtin=1}^\InnerIt$ to denote
the sequence of inner iterates within $\real^\numnode$.  For the
inner phase corresponding to outer update from
\mbox{$\mytheta{\newt} \rightarrow \mytheta{\newt+1}$,} the inner
phase takes the initialization $\myinnerest{1} \leftarrow
\mytheta{\newt}$, and then reduces as output $\myinnerest{\InnerIt}
\rightarrow \mytheta{\newt+1}$ to the outer iteration.  In more
detail, the inner phase can be broken down into three steps, which
we now describe in detail.


\paragraph{Step 1, deciding the averaging direction}  The first step
is to choose a direction in which to perform averaging.  In a single
cycle graph, since left and right are viewed as the same, there is
only one choice, and hence nothing to be decided.  In contrast, the
grid or RGG graphs require a decision-making phase, which proceeds
as follows.  One node in the first (bottom left) square, wakes up
and chooses uniformly at random to send in the horizontal or
vertical direction. We code this decision using the random variable
$\zeta\in\{-1,1\}$, where $\zeta = -1$ (respectively $\zeta = +1$)
represents the horizontal (respectively vertical) direction.  To
simplify matters, we assume in the remainder of this description
that the averaging direction is horizontal, with the modifications
required for vertical averaging being standard.

\paragraph{Step 2, choosing the head nodes}

\begin{figure}
\setlength{\unitlength}{1mm} \centering
\begin{tabular}{cc}
\begin{picture}(50,50)
\put(0,0){\dashbox{1}(50,50)}
\multiput(10,0)(10,0){4}{\dashbox{1}(0,50)}
\multiput(0,10)(0,10){4}{\dashbox{1}(50,0)}
\multiput(0,0)(0,10){5}{\multiput(5,5)(10,0){5}{\circle*{1}}}
\put(7,8){\circle{1}}\put(7,8){\line(-1,1){5}}\put(5.5,5){\mbox{$s_{11}$}}
\put(2,13){\circle{1}}\put(2,13){\line(1,2){6}}\put(0,15){\mbox{$s_{12}$}}
\put(8,25){\circle{1}}\put(8,25){\line(0,1){12}}\put(3,26){\mbox{$s_{13}$}}
\put(8,37){\circle{1}}\put(8,37){\line(-1,1){5}}\put(3,36){\mbox{$s_{14}$}}
\put(3,42){\circle{1}}\put(0,44){\mbox{$s_{15}$}}
\end{picture} & \quad \quad \quad
\begin{picture}(50,50)
\put(0,0){\dashbox{1}(50,50)}
\multiput(10,0)(10,0){4}{\dashbox{1}(0,50)}
\multiput(0,10)(0,10){4}{\dashbox{1}(50,0)}
\multiput(0,0)(0,10){5}{\multiput(5,5)(10,0){5}{\circle*{1}}}
\put(5,3){\mbox{$\Path_{1}$}}\put(7,8){\circle{1}}\put(7,8){\line(1,-1){5}}\put(12,3){\circle{1}}\put(12,3){\line(1,0){15}}\put(27,3){\circle{1}}\put(27,3){\line(1,1){5}}\put(32,8){\circle{1}}\put(32,8){\line(3,-1){15}}\put(47,3){\circle{1}}
\put(.5,15){\mbox{$\Path_2$}}\put(2,13){\circle{1}}\put(2,13){\line(1,0){15}}\put(17,13){\circle{1}}\put(17,13){\line(1,1){5}}\put(22,18){\circle{1}}\put(22,18){\line(1,0){15}}\put(37,18){\circle{1}}\put(37,18){\line(1,-1){6}}\put(43,12){\circle{1}}
\put(5,21.5){\mbox{$\Path_3$}}\put(8,25){\circle{1}}\put(8,25){\line(2,1){6}}\put(14,28){\circle{1}}\put(14,28){\line(2,-1){10}}\put(24,23){\circle{1}}\put(24,23){\line(3,1){12}}\put(36,27){\circle{1}}\put(36,27){\line(1,0){12}}\put(48,27){\circle{1}}
\put(5,31.5){\mbox{$\Path_4$}}\put(8,37){\circle{1}}\put(8,37){\line(1,0){10}}\put(18,37){\circle{1}}\put(18,37){\line(1,-1){5}}\put(23,32){\circle{1}}\put(23,32){\line(1,0){10}}\put(33,32){\circle{1}}\put(33,32){\line(2,1){10}}\put(43,37){\circle{1}}
\put(0,44){\mbox{$\Path_5$}}\put(3,42){\circle{1}}\put(3,42){\line(3,1){15}}\put(18,47){\circle{1}}\put(18,47){\line(1,0){10}}\put(28,47){\circle{1}}\put(28,47){\line(1,-1){5}}\put(33,42){\circle{1}}\put(33,42){\line(1,0){12}}\put(45,42){\circle{1}}
\end{picture}\\
(a) & \quad \quad \quad (b)\\
\end{tabular}
\caption{(a) The node labeled $s_{11}$ in the first square, chooses
the horizontal direction for averaging ($\zeta=-1$); it
  passes the token vertically to inform other nodes to average
  horizontally.  Nodes who receive the token pass it to another node
  in the above adjacent square. (b) The head nodes $s_{1j}$
  $j=1,2,\cdots$, as determined in the first step, establish routes
  horizontally ($\Path_j$, $j=1, 2, \cdots, m$) and then average along these paths.}
\label{fig:path_establishing}
\end{figure}
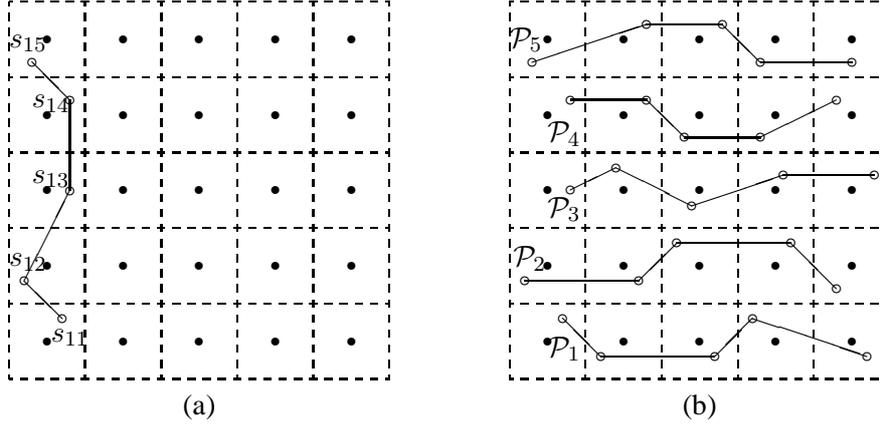

This step applies only to the grid and RGG graphs.  Given our
assumption that the node in the first square has chosen the
horizontal direction, it then passes a token message to a randomly
selected node in the above adjacent square. The purpose of this
token is to determine which node (referred to as the head node)
should be involved in establishing the route passing through the
given square.  After receiving the token, the receiving node passes
it to another randomly selected node in the above adjacent square
and so on. Note that in the special case of grid, there is only one
node in each square, and so no choices are required within squares.
After $m$ rounds, one node in each square $(x_1,y_j),
j=1,2,\cdots,m$ ($(x_i,y_1), i=1,2,\cdots,m$) receives the token, as
illustrated in Figure~\ref{fig:path_establishing}.  Note that again
in a single cycle graph, there is nothing to be decided, since the
direction and head nodes are all determined.

\paragraph{Step 3, establishing routes and averaging}
In this phase, each of head nodes establishes a horizontal path, and
then perform averaging along the path, as illustrated in
Figure~\ref{fig:path_establishing}(b). This part of algorithm
involves three substeps, which we now describe in detail.
\bcar
\item
For $j=1,2,\cdots,m$, each head node $s_{1j}$ selects a node $s_{2j}$
uniformly at random (u.a.r.) from within the right adjacent square,
and passes to it the quantity $\Emyinnerest{1j}{1}$.  Given the
Gaussian noise model, node $s_{2j}$ then receives the quantity
\begin{align*}
\tgam_{1j}(1) & = \Emyinnerest{1j}{1}+v_{1j}, \quad \mbox{where
  $v_{1j}\sim N(0,\sig)$,}
\end{align*}
and then updates its own local variable as \mbox{$\Emyinnerest{2j}{2}
= \Emyinnerest{2j}{1} + \tgam_{1j}(1)$.} We then iterate this same
procedure---that is, node $s_{2j}$ selects another $s_{3j}$
u.a.r. from its right adjacent square, and passes the message
$\Emyinnerest{2j}{2}$.  Overall, at round $i$ of this update
procedure, we have
\begin{align*}
\Emyinnerest{(i+1)j}{i+1} & = \Emyinnerest{(i+1)j}{i} + \tgam_{ij}(i),
\end{align*}
where $\tgam_{ij}(i)=\gamma_{ij}(i)+v_{ij}$, and $v_{ij}\sim
N(0,\sig)$.  At the end of round $m$, node $s_{mj}$ can compute a
noisy version of the average along the path $\Path_j: s_{1j}\to
s_{2j}\to\cdots\to s_{mj}$, in particular via the rescaled quantity
\begin{align*}
\eta_j \defn\frac{\gamma_{mj}(m)}{m}=
\frac{1}{m}\sum_{i=1}^{m}\theta_{s_{ij}}(t)+v_j \quad j = 1, 2,
\cdots, m.
\end{align*}
Here the variable $v_j\sim\mathcal{N}(0,\frac{\sig}{m})$, since the
noise variables associated with different edges are independent.


\item
At this point, for each $j = 1, 2, \ldots, m$, each node $s_{mj}$
which has the noisy version, $\eta_j$, of the path average along
route $\Path_j$; can share this information with other nodes in the
path by sending $\eta_j$ back to the head node.  A naive way to do
this is as follows: node $s_{mj}$ makes $m$ copies of
$\eta_j$---namely, $\etajl=\eta_j$, $l=1,2,\cdots,m$---and starts
transmitting one copy at a time back to the head node. Nodes along
the path simply forward what they receive, so that after $m-i+m-1$
time steps, node $s_{ij}$ receives $m$ noisy copies of the average,
$\tetaijl=\etajl+\vij$ where $\vij\sim\mathcal{N}(0,(m-i)\sig)$.
Averaging the $m$ copies, node $s_{ij}$ can compute the quantity
\begin{align*}
\Emyinnerest{ij}{3m-i-1} & \defn \frac{1}{m} \sum_{l=1}^{m}\tetaijl
\; = \; \frac{1}{m} \sum_{l=1}^{m} \bth_{s_{lj}}(\newt) + w_{ij},
\end{align*}
where $w_{ij}=v_j+\frac{1}{m}\sum_{l=1}^{m}\vij$. Since the noise on
different links and different time steps are independent Gaussian
random variables, we have $w_{ij} \sim \mathcal{N}(0,\sig_i)$,
with
\begin{align*}
\sig_i & = \frac{1}{m}\sig+(1-\frac{i}{m})\sig=(1-\frac{(i-1)}{m})\sig
\: \leq \; \sig.
\end{align*}

Therefore, at the end of $M=\Theta (m)$ rounds, for each \mbox{$j =
1, 2, \ldots, m$,} all nodes have the average of the estimates in
the path $\Path_j$ that is perturbed by Gaussian noise with variance
at most $\sig$. Since $m=\Theta(\diam)$, we have
$\InnerIt=\Theta(\diam)$.

\item

At the end of the inner phase $\newt$, nodes that were involved in a
path use their estimate of the average along the path to update
$\mytheta{\newt}$, while estimate of the nodes that were not
involved in any route remain the same.  A given node $s_{ij}$ on a
path updates its estimate via
\begin{equation}\label{eq:update_scaler}
\theta_{s_{ij}}(\newt+1)= \big \{ 1-\epsp(\newt) \big
\}\theta_{s_{ij}}(\newt)+\epsp(\newt) \gamma_{ij}(M),
\end{equation}
where $\epsp(\newt) = \bO{ \frac{1}{\newt + 1/\delta}}$.  On the
other hand, using $\inprod{\cdot}{\cdot}$ to denote the Euclidean
inner product, we have \mbox{$\gamma_{ij}(M)=
\inprod{w}{\theta(\newt)} +
 v_{s_{ij}}$,} where $w$ is the averaging vector of the route $\Path_j$ with the entries $w(s_{\ell j})=\frac{1}{m}$ for
 \mbox{$\ell=1,2,\cdots,m$,} and zero otherwise. Combining the scalar
 updates~\eqref{eq:update_scaler} yields the matrix-form update
\begin{align}
\label{EqnMatrixForm}
\bth(\newt+1) & = \bth(\newt) - \epsp(\newt) \big \{(I - \bW(\newt))
\bth(\newt) + \bvp(\newt) \},
\end{align}
where the matrix \mbox{$\bW(\newt) =\bW(\newt; \Path_1, \Path_2,
\cdots,
  \Path_m,\zeta)$} is a random averaging matrix induced by the choice
of routes $\Path_1, \Path_2, \cdots, \Path_m$ and the random
directions $\zeta$.  The noise vector $\bvp(\newt) \sim
\mathcal{N}(0,\bCp)$ is additive noise.  Note that for any given time,
the noise at different nodes are correlated via the matrix $\bCp$, but
for different time instants $\newt \neq \newt'$, the noise vectors
$\bvp(\newt)$ and $\bvp(\newt')$ are independent.  Moreover, from our
earlier arguments, we have the upper bound $\max \limits_{i=1, \ldots,
  \numnode} \Cp_{ii} \leq \sig$.

\ecar



\section{Proof of Theorem~\ref{ThmMain}}
\label{SecProof}

We now turn to the proof of Theorem~\ref{ThmMain}.  At a high-level,
the structure of the argument consists of decomposing the vector
$\bth(\newt) \in \real ^ \numnode$ into a sum of two terms: a
component within the consensus subspace (meaning all values of the
vector are identical), and a component in the orthogonal complement.
Using this decomposition, the mean-squared error splits into a sum
of two terms and we use standard techniques to bound them. As will
be shown, these bounds depend on the parameter $\delta$, noise
variance, the initial MSE, and finally the (inverse) spectral gap of
the update matrix. The final step is to lower bound the spectral gap
of our update matrix.


\subsection{Setting up the proof} \label{SubSecSetUp}

Recalling the averaging matrix $\bW(\newt)$ from the
update~\eqref{EqnMatrixForm}, we define the Laplacian matrix
\mbox{$\SpecLap(\newt) \defn I -\bW(\newt)$.}  We then define the
average matrix $\bbW \defn \Exp{\bW(\newt)}$, where the expectation is
taken place over the randomness due to the choice of routes;\footnote{
  For the single cycle graph, there is only one route that involves
  all the nodes at each round, so $\bW(\newt)$ is deterministic in
  this case.}  in a similar way, we define the associated (average)
Laplacian $\SpecLapAve \defn I - \bbW$.  Finally, we define the
rescaled quantities
\begin{align}\label{EqnRescaled}
\eps(\newt) \defn \lam_2(\SpecLapAve) \: \epsp(\newt), \quad
\bL(\newt) \defn \frac{1}{\lam_2(\SpecLapAve)} \: \SpecLap(\newt),
\quad
\mbox{and} \quad \bv(\newt)  \defn \frac{1}{\lam_2(\SpecLapAve)} \:
v'(\newt),
\end{align}
where we recall that $\lam_2(\cdot)$ denotes the second smallest
eigenvalue of a symmetric matrix.  In terms of these rescaled
quantities, our algorithm has the form
\begin{equation}
\label{EqnScaledUpdate}
\bth(\newt+1) =
\bth(\newt)-\eps(\newt) [\bL(\newt) \bth(\newt) + \bv(\newt)],
\end{equation}
as stated previously in the update equation~\eqref{eq:outer_phase}.
Moreover, by construction, we have \mbox{$\bv(\newt) \sim
  \mathcal{N}(0,\bC)$} where
\mbox{$\bC=\frac{1}{(\lam_2(\bar{\SpecLap}))^2}\bCp$}. We also, for
theoretical convenience, set
\begin{align}\label{EqnEpsValue}
\epsp (\newt) = \frac{1}{\SpecGap (\newt + \invdelt)},
\end{align}
or equivalently $\eps (\newt) = \frac{1}{(\newt + \invdelt)}$ for
$\newt = 1, 2, \cdots$.

We first claim that the matrix $\bbW$ is symmetric and (doubly)
stochastic.  The symmetry follows from the fact that different routes
do not collide, whereas the matrix is stochastic because every row of
$\bW$ (depending on whether the node corresponding to that row
participates in a route or not) either represents an averaging along a
route or is the corresponding row of the identity matrix.
Consequently, we can interpret $\bbW$ as the transition matrix of a
reversible Markov chain.  It is an irreducible Markov chain, because
within any updating round, there is a positive chance of averaging
nodes that are in the same column or row, which implies that the
associated Markov chain can transition from one state to any other in
at most two steps. Moreover, the stationary distribution of the chain
is uniform (i.e., $\pi = \bone/n$).

We now use these properties to simplify our study of the sequence
$\{\bth(\newt)\}_{\newt=1}^{\infty}$ generated by the update
equation~\eqref{EqnScaledUpdate}. Since $\SpecLapAve$ is real and
symmetric, it has the eigenvalue decomposition $\SpecLapAve = \bU
\bLam \bU^T$, where $\bU =
\begin{bmatrix} \bu_1 & \bu_2 & \cdots & \bu_n
\end{bmatrix}$ is a unitary matrix (that is, $\bU^T \bU = \bI_n$).
Moreover, we have $\bLam = \diag \{ \lam_1(\SpecLapAve),
\lam_2(\SpecLapAve), \cdots, \lam_n(\SpecLapAve)\}$, where
$\lam_i(\SpecLapAve)$ is the eigenvalue corresponding to the
eigenvector $\bu_i$, for $i = 1, \ldots, n$.  Since $\LapAve =
\frac{1}{\lam_2(\SpecLapAve)}(\bI-\bbW)$, the eigenvalues of
$\LapAve$ and $\bbW$ are related via
\begin{eqnarray*}
\lam_i(\LapAve) & = &
\frac{1}{\lam_2(\SpecLapAve)}(1-\lam_{n+1-i}(\bbW)) \\
& = & \frac{1}{1-\lam_{n-1}(\bbW)}(1-\lam_{n+1-i}(\bbW)).
\end{eqnarray*}
Since the largest eigenvalue of an
irreducible Markov chain is one (with multiplicity
one)~\cite{Grimmett}, we have $1 = \lam_n(\bbW) > \lam_{n-1}(\bbW)
\geq \cdots \geq \lam_1(\bbW)$, or equivalently
\begin{equation*}
0 = \lam_1(\LapAve) < \lam_2(\bbL) \leq \cdots \leq \lam_n(\bbL),
\end{equation*}
with $\lam_2 (\bbL) = 1$. Moreover, we have $\SpecLapAve \bone =
\LapAve \bone = \bzero$, so that the first eigenvector $u_1 =
\bone/\sqrt{n}$ corresponds to the eigenvalue $\lam_1(\LapAve) = 0$.
Let $\tbU$ denote the matrix obtained from $\bU$ by deleting its
first column, $\bu_1$. Since the smallest eigenvalue of $\LapAve$ is
zero, we may write $\LapAve = \tbU \tbLam \tbU^T$, where
$\tbLam=\diag\{\lam_2(\bbL),\cdots\lam_n(\bbL)\}$,
$\tbU^T\tbU=\bI_{n-1}$, and $\tbU\tbU^T=\bI_{n} -
\frac{\bone\bone^T}{n}$.
With this notation, our analysis is based on the decomposition
\begin{align}
\label{EqnDecTheta}
\bth(\newt) & = \alp(\newt)\frac{\bone}{\sqrt{n}}+\tbU\bbet(\newt),
\end{align}
where we have defined \mbox{$\alp(\newt) \defn
\inprod{\bone/\sqrt{n}}{\bth(\newt)} \in \real$} and
\mbox{$\bbet(\newt) \defn \tbU^T \bth(\newt) \in \real^{n-1}$.}
Since $\bone^T \bL (\newt) = \bzero^T$ for all $\newt = 1, 2,
\cdots$, from the decomposition~\eqref{EqnDecTheta} and the form of
the updates~\eqref{EqnScaledUpdate}, we have the following
recursions,
\begin{align}
\label{EqnAlphaRecur} \alp(\newt+1) & = \alp(\newt) - \eps(\newt)
\frac{\bone^T}{\sqrt{n}} \bv(\newt), \quad\text{and}
\end{align}
\begin{align}
\label{EqnBetaRecur} \beta(\newt+1) & = \beta(\newt) - \eps(\newt)
\big( \Ltil(\newt) \beta(\newt) + \tbU^T \bv(\newt) \big).
\end{align}
Here $\Ltil$ is an $(\numnode-1)\times(\numnode-1)$ matrix defined by
the relation
\begin{align*}
\bU^T \bL (\newt) \bU  & =
\begin{bmatrix}
0 & \bzero^T \\ \bzero & \Ltil(\newt) \\
\end{bmatrix}_{\numnode\times \numnode}.
\end{align*}


\subsection{Main steps}

As we show, part (a) of the theorem requires some intermediate
results of the proof of part (b). Accordingly, we defer it to the
end of the section. With this set-up, we now state the two main
technical lemmas that form the core of Theorem~\ref{ThmMain}. Our
first lemma concerns the behavior of the component sequences
$\{\alp(\newt)\} _ {\newt = 0} ^ {\infty}$ and $\{ \bbet (\newt) \}
_ {\newt=0} ^ {\infty}$ which evolve according to equations
\eqref{EqnAlphaRecur} and \eqref{EqnBetaRecur} respectively.
\begin{lemma}
\label{LemMSE} Given the random sequence $\{\bth(\newt)\}$ generated
by the update equation \eqref{eq:outer_phase}, we have
\begin{align}
\label{EqnMSEDecomp}
\MSE(\bth(\newt)) & = \underbrace{\frac{1}{n}
  \var{\alp(\newt)}}_{e_1(\newt)} + \underbrace{\frac{1}{n} \;
  \Exs[\|\bbet(\newt)\|_2^2]}_{e_2(\newt)}.
\end{align}
Furthermore, $e_1 (\newt)$ and $e_2(\newt)$ satisfy the following
bounds:
\begin{enumerate}
\item[(a)] For each iteration $\newt = 1,2, \ldots$, we have
\begin{align}
\label{EqnEoneBound} e_1(\newt) & \leq \frac{ \sigma^2 \, \delta}
{[\SpecGap] ^2}.
\end{align}

\item[(b)]Moreover, for each iteration $\newt = 1,2, \ldots$ we have
\begin{align} \label{EqnEtwoBound}
\Serr (\newt) \, & \le \, \frac {\sig} {[\SpecGap]^2} \: \frac{\log
(\newt + \invdelt - 1)} {\newt + \invdelt - 1} \, + \, \Serr (0) \:
\frac{\invdelt - 1} {\newt + \invdelt - 1},
\end{align}
\end{enumerate}
\end{lemma}
%


From Lemma~\ref{LemMSE}, we conclude that in order to guarantee an
$\order( \frac {\sig \delta} {[\SpecGap] ^2})$ bound on the MSE, it
suffices to take $\newt$ such that
\begin{align*}
\frac {\invdelt - 1} {\newt + \invdelt - 1} \, \le \, \frac{\sig \:
\delta} {\Serr (0) [\SpecGap] ^2}, \quad \text{and} \quad
\frac{\log (\newt + \invdelt - 1)} {\newt + \invdelt - 1} \, \le \,
\delta.
\end{align*}
Note that the first inequality is satisfied when $\newt \ge
\frac{\Serr(0)}{\sig \delta ^2} \: [\SpecGap] ^2$. Moreover, doing a
little bit of algebra, one can see that $\newt = \frac{2}{\delta}
\log \frac{1}{\delta} - (\invdelt - 1)$ is sufficient to satisfy the
second inequality. Accordingly, we take
\begin{align*}
\newt = \MAX {\frac{2} {\delta} \log{\invdelt}} {\frac{\Serr(0) [\SpecGap] ^2}{\sig \delta ^2}}
\end{align*}
outer iterations.

The last part of the proof is to bound the second smallest
eigenvalue of the Laplacian matrix $\SpecLapAve$. The following
lemma, which we prove in Section~\ref{SecLemEigenvalue} to follow,
addresses this issue. Recall that $\lam_2(\cdot)$ denotes the second
smallest eigenvalue of a matrix.
\begin{lemma}
\label{LemEigenvalue}
The averaged matrix $\SpecLapAve$ that arises from our protocol has
the following properties:
\begin{enumerate}[(a)]
\item For a cycle and a regular grid we have $\SpecGap = \Omega(1)$, and
\item for a random geometric graph, we have \mbox{$\SpecGap = \Omega ( \frac{1}{\log{n}})$}, with high
probability.
\end{enumerate}
\end{lemma}

\noindent It is important to note that the averaged matrix
$\SpecLapAve$ is \emph{not the same} as the graph Laplacian that
would arise from standard averaging on these graphs.  Rather, as a
consequence of establishing many paths and averaging along them in
each inner phase, our protocol ensures that the matrix behaves
essentially like the graph Laplacian for the fully connected graph.

As established previously, each outer step requires $\InnerIt =
\order(\diam)$ iterations. Therefore, we have shown that it is
sufficient to take a total of
\begin{align*}
\Tfinal \, = \, \bO{ \diam \: \MAX {\frac{2} {\delta}
\log{\invdelt}} {\frac{\Serr(0) [\SpecGap] ^2}{\sig \delta ^2}} }
\end{align*}
transmissions per edge in order to guarantee a $ \: \frac {3 \: \sig
\delta} {[\SpecGap] ^2}$ bound on the MSE. As we will see in the
next section, assuming that the initial values are fixed, we have
$\Ferr (0) = 0$, hence $\MSE (\bth (0)) = \Serr (0)$. The claims in
Theorem~\ref{ThmMain} then follow by standard calculations of the
diameters of the various graphs and the result of the Lemma
\ref{LemEigenvalue}.

It remains to prove the two technical results, Lemma~\ref{LemMSE}
and~\ref{LemEigenvalue}, and we do so in the following sections.


\subsection{Proof of Lemma~\ref{LemMSE}}
\label{SecLemMSE}

We begin by observing that
\begin{align*}
\Exp{(\bth(\newt)-\bath\bone) (\bth(\newt)-\bath\bone)^T} & = \Term_1
+ \Term_2 + \Term_3,
\end{align*}
where $\Term_1 \defn \Exp{(\alp(\newt) - \sqrt{n}\bath)^2}
\frac{\bone\bone^T}{\numnode}$, the second term is given by
$\Term_2  \defn \Exp{\tbU\bbet(\newt)\bbet(\newt)^T\tbU^T}$,
and
\begin{align*}
\Term_3  \defn \Exp{ (\alp(\newt) - \sqrt{n}\bath ) \:
\frac{\bone}{\sqrt{n}} \: \bbet(\newt)^T \tbU^T} \; + \;
\Exp{(\alp(\newt)-\sqrt{n}\bath) \: \tbU \bbet(\newt) \:
\frac{\bone^T}{\sqrt{n}}}.
\end{align*}
Since $\tbU$ has orthonormal columns, all orthogonal to the all one
vector ($\bone^T\tbU=\bzero$), it follows that
\mbox{$\trace(\Term_2) = \Exs \big[ \|  \bbet(\newt) \|_2^2]$,} and
\mbox{$\trace(\Term_3) = 0$.}

It remains to compute $\trace(\Term_1)$. Unwrapping the
recursion~\eqref{EqnAlphaRecur} and using the fact that initialization
$\bth(0)$ implies $\alp(0) = \sqrt{n} \thetacent$ yields
\begin{align}
\label{EqnUnrappedAlpha} \alp(\newt) & = \sqrt{n} \thetacent -
\sum_{l=0}^{\newt-1}\eps(l) \,
\inprod{\frac{\bone}{\sqrt{n}}}{\bv(l)},
\end{align}
for all $\newt = 1, 2, \ldots$. Since $\bv(l)$, $l = 0, 1, \cdots,
\newt-1$, are zero mean random vectors, from equation
\eqref{EqnUnrappedAlpha} we conclude that $\Exs[\alp(\newt)] =
\sqrt{n} \bath$ \footnote{Here we have assumed that the initial
values, $\bth_i (0)$ $i = 1, 2, \cdots, \numnode$, are given
(fixed).} and accordingly, $\trace(\Term_1)=\var{\alp(\newt)}$.
Recalling the definition of the MSE \eqref{EqnMSEDefinition} and
combining the pieces yields the
claim \eqref{EqnMSEDecomp}. \\

\noindent (a) From equation \eqref{EqnUnrappedAlpha}, it is clear
that each $\alp(\newt)$ is Gaussian with mean $\sqrt{n} \thetacent$.
It remains to bound the variance. Using the i.i.d. nature of the
sequence $\bv(i) \sim \mathcal{N}(0,\bC)$, we have
\begin{align*}
\var{\alp(\newt)} & = \Exs \biggr[ \big(\sum_{l=0}^{\newt-1} \eps(l)
  \inprod{\frac{\bone}{\sqrt{n}}}{ \bv(l)} \big)^2 \biggr] \\
& = \sum_{l=0}^{\newt-1} \frac{\eps(l)^2}{n} \inprod{\bone}{\bC \bone} \\
& = \sum_{l=0}^{\newt-1}\epsp(l)^2 \frac{\inprod{\bone}{\bCp
\bone}}{n},
\end{align*}
where we have recalled the rescaled quantities~\eqref{EqnRescaled}.
Recalling the fact that $\Cp_{ii} \leq \sigma^2$ and using the
Cauchy-Schwarz inequality, we have $\Cp_{ij} \leq
\sqrt{\Cp_{ii}\Cp_{jj}} \; \leq \; \sigma^2$. Hence, we obtain
\begin{align*}
\var{\alp(\newt)} & \leq n \sigma^2 \; \sum_{l = 0 } ^ {\newt-1}
\epsp(l)^2
\\
& = \frac {n \sig} {[\SpecGap] ^2}
\sum_{l = 0} ^ {\newt-1} \frac{1}{(\frac{1} {\delta} + l )^2} \\
& \leq \frac {n \sig} {[\SpecGap] ^2}
\int_{\frac{1}{\delta}}^{\infty}\frac{1}{x^2}dx\; = \frac {n \; \sig
\delta} {[\SpecGap] ^2};
\end{align*}
from which rescaling by $1/n$ establishes the
bound~\eqref{EqnEoneBound}. \\

\noindent (b) Defining $\VF (\bbet(\newt), \bv (\newt)) = \Ltil
(\newt) \bbet (\newt) + \tbU^T \bv (\newt)$, the update equation
\eqref{EqnBetaRecur} can be written as
\begin{align*}
\bbet (\newt + 1) \, = \, \bbet (\newt) \, - \, \eps \: (\newt)
\VF(\bbet(\newt), \bv (\newt)),
\end{align*}
for $\newt = 1, 2, \cdots$. In order to upper bound $\Serr (\newt +
1)$, defined in \eqref{EqnMSEDecomp}, we need to control $\Serr
(\newt + 1) - \Serr (\newt)$. Doing some algebra yields
\begin{align*}
\Serr (\newt + 1) - \Serr (\newt) \, & = \, \frac{1}{\numnode} \:
\Exp{ \inprod {\bbet (\newt + 1) -
\bbet (\newt)} {\bbet (\newt + 1) + \bbet (\newt)} } \\
\, & = \, \frac{1}{\numnode} \: \Exp{ \inprod {- \eps \: (\newt) \VF
(\bbet (\newt, \bv (\newt)))} {- \eps \: (\newt) \VF (\bbet (\newt,
\bv (\newt)))  + 2 \bbet(\newt)}},
\end{align*}
and hence
\begin{align*}
\Serr (\newt + 1) - \Serr (\newt) \,  = \, \frac{1}{\numnode} \:
\eps(\newt)^2 \: \Exp {\vnorm{\VF (\bbet (\newt), \bv(\newt))}^2}
\,- \, \frac{2\eps(\newt)}{\numnode} \: \Exp{\inprod {\VF (\bbet
(\newt), \bv(\newt))} {\bbet (\newt)}}.
\end{align*}
Since $\bbet (\newt)$ is independent of both $\Lap (\newt)$ and $\bv
(\newt)$, by conditioning on the $\bbet(\newt)$ and using the tower
property of expectation, we obtain
\begin{align*}
\Exp{\inprod {\VF (\bbet (\newt), \bv(\newt))} {\bbet (\newt)}} \, &
= \, \Exp{\inprod {\Exp{\Ltil} \bbet (\newt)} {\bbet (\newt)}}.
\end{align*}
By construction all the eigenvalues of $\Exp{\Ltil}$ are greater
than one, hence
\begin{align*}
\inprod {\Exp{\Ltil} \bbet (\newt)} {\bbet (\newt)} \; \ge \;
\vnorm{\bbet (\newt)} ^2.
\end{align*}
Putting the pieces together, we obtain
\begin{align}
\nonumber \Serr (\newt + 1) \, & \le \, \frac{1}{\numnode} \: \eps
(\newt)^2 \: \Exp {\vnorm {\VF (\bbet(\newt), \bv(\newt))}^2} \, +
\, (1 - 2 \eps
(\newt)) \: \Serr (\newt) \\
\label{EqnFirstBoundEtwo} \, &= \, \, \frac{1}{\numnode} \: \eps
(\newt)^2 \: \underbrace{\Exp {\vnorm {\Ltil (\newt) \bbet (\newt)}
^2}} _{\Term_1} \, + \, \frac{1}{\numnode} \: \eps (\newt)^2 \:
\underbrace{\Exp {\vnorm {\tbU^T \bv (\newt)} ^2}} _{\Term_2} \, +
\, (1 - 2 \eps (\newt)) \: \Serr (\newt),
\end{align}
where we used the fact that \mbox{$\Exp {\inprod {\Ltil (\newt)
\bbet (\newt)} {\tbU^T \bv (\newt)}} = 0$}. We continue by upper
bounding the terms \mbox{$\Term_1 = \Exp {\vnorm {\Ltil (\newt)
\bbet (\newt)} ^2}$}, and $\Term_2 = \Exp {\vnorm {\tbU^T \bv
(\newt)} ^2}$. First, we bound the former. By definition of the
$l_2$-operator norm, we have
\begin{align*}
\Exp {\vnorm {\Ltil (\newt) \bbet (\newt)} ^2} \, & \le \,
\Exp{\matsnorm{\Ltil (\newt)}{2} ^2 \: \vnorm {\bbet (\newt)} ^2}.
\end{align*}
On the other hand, using the fact that $\Ltil (\newt) = \frac{1}
{\SpecGap} \: \tbU^T \: (\bI - \bW (\newt)) \: \tbU$ (recall the
identities of the Section \ref{SubSecSetUp}) yields\footnote{Let $v$
be an eigenvector of the matrix $\bW(\newt)$ corresponding to the
eigenvalue $\lam \neq 1$. Since $\bone ^T v = 0$, there exist an
$(n-1)$-dimensional vector $u$ such that $v = \tbU u$. Therefore we
have,
\begin{align*}
\tbU^T (\bI - \bW(\newt)) \tbU u \, = \, \tbU^T (\bI - \bW(\newt)) v
\, = \, (1 - \lam) \tbU^T v \, = \, (1 - \lam) u.
\end{align*}
So by subtracting one from the eigenvalues of $\tbU^T (\bI -
\bW(\newt)) \tbU$, we obtain the non-one eigenvalues of
$\bW(\newt).$}
\begin{align*}
\matsnorm{\Ltil (\newt)}{2} \, & \le \, \frac{1}{\SpecGap} \: (1 +
\matsnorm {\bW(\newt)}{2}) \, = \, \frac {2}{\SpecGap}.
\end{align*}
Therefore, we have the following bound on $\Term_1$
\begin{align}\label{EqnBoundTermOne}
\Term_1 \, \le \, \frac{4}{[\SpecGap] ^2} \: \Exp{\vnorm {\bbet
(\newt)} ^2}.
\end{align}
Turning to term $\Term_2$, we have
\begin{align}\label{EqnBoundTermTwo}
\Term_2 \,  = \, \Exp {\bv (\newt)^T (\bI - \frac{\bone
\bone^T}{\numnode}) \bv (\newt) } \, & \le \, \trace \big( \cov
(\bv(\newt)) \big) \, \le \, \frac {\numnode \sig} {[\SpecGap] ^2}.
\end{align}
Substituting the inequalities \eqref{EqnBoundTermOne} and
\eqref{EqnBoundTermTwo} into \eqref{EqnFirstBoundEtwo}, we obtain
the following recursive bound on $\Serr (\newt + 1)$
\begin{align*}
\Serr (\newt + 1) \, & \le \, \frac{\sig}{[\SpecGap] ^2} \: \eps
(\newt) ^2 \, + \, \left(1 - 2\eps (\newt) + \frac {4 \eps(\newt)
^2}{[\SpecGap] ^2}\right) \: \Serr (\newt).
\end{align*}
Recall the definitions \eqref{EqnRescaled} and \eqref{EqnEpsValue}.
If $\delta \le \frac{[\SpecGap] ^2}{4}$, then $1 - 2\eps (\newt) +
\frac {4 \eps(\newt) ^2}{[\SpecGap] ^2} \le 1 - \eps (\newt)$, and
hence we have
\begin{align}\label{EqnFirstRecurBoundEtwo}
\Serr (\newt + 1) \, &\le \, \frac{\sig}{[\SpecGap] ^2} \eps (\newt)
^2 \, + \, (1 - \eps (\newt)) \Serr (\newt),
\end{align}
for all $\newt = 1, 2, \cdots$. Unwrapping the inequality
\eqref{EqnFirstRecurBoundEtwo} yields
\begin{align}\label{EqnSecoRecurBoundEtwo}
\Serr (\newt + 1) \, & \le \, \frac{\sig}{[\SpecGap] ^2} \: \sum _{k
= 0} ^{\newt} \eps (k)^2 \prod _{l = k + 1} ^{\newt} (1 - \eps (l))
\, + \, \prod _{l = 0} ^{\newt} (1 - \eps (l)) \: \Serr (0).
\end{align}
On the other hand, the product $\prod _{l = k+1} ^{\newt} (1 - \eps
(l))$ forms a telescopic series and is equal to $\frac{k +
\invdelt}{\newt + \invdelt}$. Substituting this fact into the
equation \eqref{EqnSecoRecurBoundEtwo} yields
\begin{align*}
\Serr (\newt + 1) \, & \le \, \frac {\sig} {[\SpecGap]^2} \: \sum
_{k=0} ^{\newt} \frac {1} {(k + \invdelt) \: (\newt + \invdelt)} \,
+ \, \Serr (0) \: \frac{\invdelt - 1} {\newt + \invdelt} \\
\, & \stackrel {(a)} {\le} \, \frac {\sig} {[\SpecGap]^2} \:
\frac{\log (\newt + \invdelt)} {\newt + \invdelt} \, + \, \Serr (0)
\: \frac{\invdelt - 1} {\newt + \invdelt},
\end{align*}
where step (a) uses the following inequality
\begin{align*}
\sum _{k=0} ^{\newt} \frac{1}{k + \invdelt} \, \le \, \int
_{\invdelt - 1} ^{\newt + \invdelt} \frac{1}{x} dx \, \le \,
\log(\newt + \invdelt),
\end{align*}
valid for $\delta \in (0, \frac{1}{2})$.


\subsection{Proof of Lemma~\ref{LemEigenvalue}}
\label{SecLemEigenvalue}

In the case of cycle there is only one averaging path and all the
nodes are involved in that at each round so the averaging matrix,
$\bW$, is fixed. More precisely, we have \mbox{$\bbW = \bW =
\frac{1}{n} \bone \bone^T $}. Therefore, $\bbW$ is a rank 1 matrix
with $\lam_{n-1}(\bbW)=0$ and accordingly we have
$\lam_{2}(\SpecLapAve) = 1 - \lam_{n-1} (\bbW) = 1$.

For the case of grid or random geometric graphs, we use the Poincare
inequality \cite{Diaconis91}. A version of this theorem can be
stated as follows: Let $\TransMat=[\entry_{ij}]$ denote the
transition matrix of an irreducible aperiodic time reversible Markov
chain with stationary distribution $\StaDist$. For each ordered pair
of nodes $(\FirNode,\SecNode)$ in the transition diagram, choose one
and only one path $\Mpath _ { \FirNode  \SecNode } = (\FirNode ,
\FirNode_1 , \FirNode_2 , \cdots , \FirNode_l , \SecNode)$ between
$\FirNode$ and $\SecNode$ and define
\begin{equation}\label{EqnPoinPathWeight}
| \Mpath _ { \FirNode  \SecNode } |
\defn \frac{1} {\StaDist (\FirNode) \entry_{\FirNode\FirNode_1}} + \frac{1} {\StaDist (\FirNode_1) \entry_{\FirNode_1 \FirNode_2}}
+ \cdots + \frac{1} {\StaDist (\FirNode_l)
\entry_{\FirNode_l\SecNode}}.
\end{equation}
Then the Poincare coefficient is
\begin{equation}\label{EqnPoinCoef}
\PoinCoef \defn \max _ { e \in \NewEdges} \sum _ { \Mpath _ {
\FirNode \SecNode } \ni e} | \Mpath _ { \FirNode  \SecNode }|
\StaDist (\FirNode) \StaDist (\SecNode),
\end{equation}
where $\NewEdges$ is the set of directed edges formed in the
previous step. Defining this quantity, the theorem states that $
\lam _ {n-1} ( \TransMat ) \le 1 - \frac{1} {\PoinCoef} $ or
equivalently,
\begin{equation}\label{EqnPoinThm}
1 - \lam _ {n-1} (\TransMat) \ge \frac {1} {\PoinCoef}.
\end{equation}
We apply this theorem to the Markov chain formed by $\bbW$; the idea
is to upper bound its Poincare coefficient.

\subsubsection{Grid}

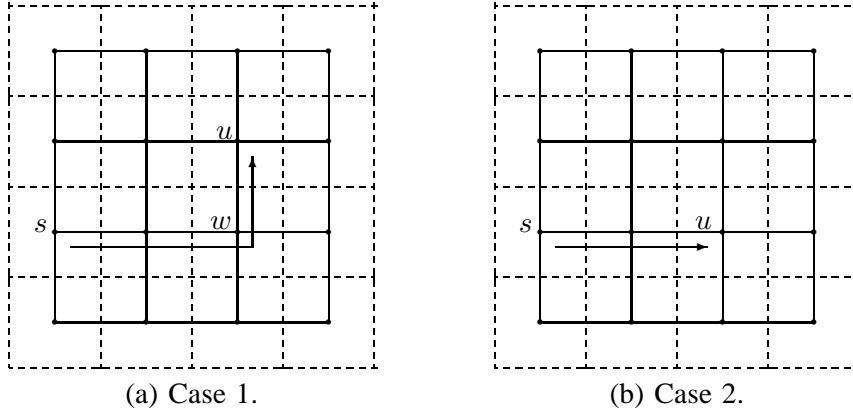
\begin{figure*}
\begin{center}
\setlength{\unitlength}{.8mm}\centering
\begin{tabular}{cc}
\begin{picture}(60,60)
\put(0,0){\dashbox{1}(60,60)}
\multiput(15,0)(15,0){4}{\dashbox{1}(0,60)}
\multiput(0,15)(0,15){4}{\dashbox{1}(60,0)}
\multiput(0,0)(0,15){4}{\multiput(7.5,7.5)(15,0){4}{\circle*{1}}}
\multiput(7.5,7.5)(0,15){4}{\line(1,0){45}}
\multiput(7.5,7.5)(15,0){4}{\line(0,1){45}}
\put(4, 22.5){\mbox{$\FirNode$}} \put(33, 23){\mbox{$\ThrNode$}}
\put(34, 38){\mbox{$\SecNode$}} \put(10, 20) {\line(1, 0){30}}
\put(40, 20) {\vector(0, 1){15}}
\end{picture} &  \quad \quad \quad
\begin{picture}(60,60)
\put(0,0){\dashbox{1}(60,60)}
\multiput(15,0)(15,0){4}{\dashbox{1}(0,60)}
\multiput(0,15)(0,15){4}{\dashbox{1}(60,0)}
\multiput(0,0)(0,15){4}{\multiput(7.5,7.5)(15,0){4}{\circle*{1}}}
\multiput(7.5,7.5)(0,15){4}{\line(1,0){45}}
\multiput(7.5,7.5)(15,0){4}{\line(0,1){45}}
\put(4, 22.5){\mbox{$\FirNode$}} \put(33, 23){\mbox{$\SecNode$}}
\put(10, 20) {\vector(1, 0){25}}
\end{picture}\\
(a) Case 1.  & \quad \quad \quad  (b) Case 2. \\
\end{tabular}
\caption{Illustration of the path $\Mpath_{\FirNode \SecNode}$ for a
grid-structured graph. (a) Case 1, where nodes $\FirNode$ and
$\SecNode$ do not belong to the same column or row. (b) Case 2,
where nodes $\FirNode$ and $\SecNode$ belong to the same column or
row. This choice of $\Mpath_{\FirNode\SecNode}$ yield a tight upper
bound on the Poincare coefficient.}
  \label{FigPathsGrid}

\end{center}
\end{figure*}

We first define a path $\Mpath _ {\FirNode \SecNode}$ for every pair
of nodes $\{\FirNode , \SecNode\}$. Two different cases can be
distinguished here. For an illustration of the path
$\Mpath_{\FirNode\SecNode}$ see Figure \ref{FigPathsGrid}.

\paragraph{Case 1} Nodes $\FirNode$ and $\SecNode$ do not
belong to the same column or row. In this case, we consider a
two-hop path \mbox{$\Mpath _ {\FirNode\SecNode} = (\FirNode \to
\ThrNode \to \SecNode)$}, where $\ThrNode = (x_{\SecNode},
y_{\FirNode})$ is the vertex of the rectangle constructed by
$\FirNode$ and $\SecNode$. Note that $x_\SecNode$ is the
$x$-coordinate of $\SecNode$ and $y_\FirNode$ is the $y$-coordinate
of $\FirNode$.  Since nodes \{$\FirNode$, $\ThrNode$\} and
\{$\ThrNode$, $\SecNode$\} are averaged $\frac{1}{2}$ of the time,
we have $ \bbW _ {\FirNode \ThrNode} = \bbW _ {\ThrNode \SecNode} =
\frac {1} {2m} $. Substituting this into \eqref{EqnPoinPathWeight}
and using the fact that $\StaDist = \frac {1} {\numnode} \bone$
yields
\begin{equation*}
| \Mpath _ {\FirNode \SecNode} | = \frac{1} {\bbW _ {\FirNode
\ThrNode} \StaDist (\FirNode)} +   \frac{1} {\bbW _ {\ThrNode
\SecNode} \StaDist (\ThrNode)} = 4m\numnode.
\end{equation*}

\paragraph{Case 2} Nodes $\FirNode$ and $\SecNode$ belong to
the same row or column. In this case, we set $\Mpath _ {\FirNode
\SecNode} = (\FirNode \to \SecNode)$ which leads to
\begin{equation*}
| \Mpath _ {\FirNode \SecNode} | = \frac {1} {\bbW _ {\FirNode
\SecNode} \StaDist (\FirNode)} = 2m\numnode.
\end{equation*}
Moreover, a given edge $e = (\FirNode \to \ThrNode)$ is involved in
at most $m$ paths. As node $\SecNode$ varies in the corresponding
column or row, we obtain $m-1$ paths in case 1, and one path in case
2.

Combining the pieces, we compute the Poincare coefficient
\begin{align*}
\PoinCoef & = \max_{e \in \NewEdges} \sum _ {\Mpath_{\FirNode
\SecNode} \ni e } | \Mpath _ {\FirNode \SecNode} | \StaDist
(\FirNode) \StaDist (\SecNode) \leq m \frac{4m\numnode} {\numnode^2}
= 4.
\end{align*}
Finally, from equation \eqref{EqnPoinThm}, we have
\begin{equation*}
\lam_2 (\SpecLapAve) = 1 - \lam_{n-1} (\bbW) \ge \frac{1}
{\PoinCoef} \ge \frac{1}{4}
\end{equation*}
which concludes the proof for the case of a grid-structured graph.

\subsubsection{Random geometric graph}

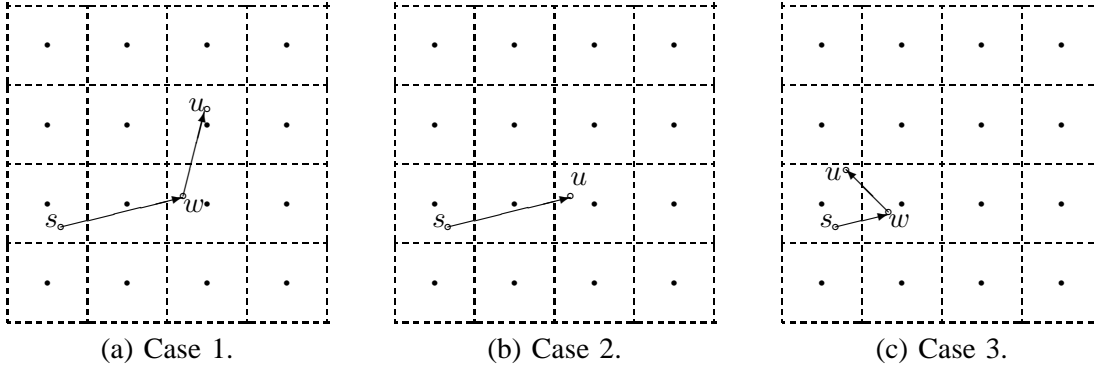
\begin{figure*}
\setlength{\unitlength}{.7mm} \centering
\begin{tabular}{cccccc}
\begin{picture}(60,60)
\put(0,0){\dashbox{1}(60,60)}
\multiput(15,0)(15,0){4}{\dashbox{1}(0,60)}
\multiput(0,15)(0,15){4}{\dashbox{1}(60,0)}
\multiput(0,0)(0,15){4}{\multiput(7.5,7.5)(15,0){4}{\circle*{1}}}
\put(10, 18){\circle{1}}\put(7, 18) {\mbox{$\FirNode$}}
\put(33, 24){\circle{1}} \put(33, 21){\mbox{$\ThrNode$}} \put(10,
18) {\vector(4, 1){23}} \put(33, 24){\vector(1, 4){4}} \put(37.5,
40.5){\circle{1}} \put(34, 40) {\mbox{$\SecNode$}}
\end{picture} &  \hspace*{.02in} &
\begin{picture}(60,60)
\put(0,0){\dashbox{1}(60,60)}
\multiput(15,0)(15,0){4}{\dashbox{1}(0,60)}
\multiput(0,15)(0,15){4}{\dashbox{1}(60,0)}
\multiput(0,0)(0,15){4}{\multiput(7.5,7.5)(15,0){4}{\circle*{1}}}
\put(10, 18){\circle{1}}\put(7, 18) {\mbox{$\FirNode$}}
\put(33, 24){\circle{1}} \put(33, 26){\mbox{$\SecNode$}} \put(10,
18) {\vector(4, 1){23}}
\end{picture} &  \hspace*{.02in} &
\begin{picture}(60,60)
\put(0,0){\dashbox{1}(60,60)}
\multiput(15,0)(15,0){4}{\dashbox{1}(0,60)}
\multiput(0,15)(0,15){4}{\dashbox{1}(60,0)}
\multiput(0,0)(0,15){4}{\multiput(7.5,7.5)(15,0){4}{\circle*{1}}}
\put(10, 18){\circle{1}}\put(7, 18) {\mbox{$\FirNode$}}
\put(10, 18) {\vector(4, 1){10}} \put(20, 21){\circle{1}} \put(20,
18){\mbox{$\ThrNode$}} \put(20, 21){\vector(-1, 1){8}} \put(12,
29){\circle{1}} \put(8, 27){\mbox{$\SecNode$}}
\end{picture} &  \hspace*{.02in} \\
(a) Case 1.  & & (b) Case 2. & & (c) Case 3. \\
& & & &
\end{tabular}
\caption{Illustration of the path $\Mpath_{\FirNode\SecNode}$ for
the case of RGG. (a) Case 1, where nodes $\FirNode$ and $\SecNode$
belong to the sub-squares in different row and columns
  (b) Case 2, where nodes $\FirNode$ and $\SecNode$
belong to the sub-squares in the same row or column. (c) Case 3,
nodes $\FirNode$ and $\SecNode$ belong to the same square. }
  \label{FigPathRGG}
\end{figure*}

For the RGG, we follow the same proof structure: namely, we first
find a path for each pair of nodes $\{\FirNode , \SecNode\}$, and
then upper bound the Poincare coefficient for the Markov chain
$\bbW$.  We first introduce some useful notation.  Let $\SquaCent :
\vertex \to \{1,2,\cdots,m\}^2$ be the mapping that takes a node as
its input and returns the sub-square of that node.  More precisely,
for some $ \FirNode \in \vertex$ we have
\begin{equation*}
\SquaCent (\FirNode) = (i,j) \quad \text{if} \, \FirNode \in
(i,j)\text{-th square} \;  i , j = 1 , 2 , \cdots, m.
\end{equation*}
Furthermore, we enumerate the nodes in square $\SquaCent (\FirNode)
= (i,j)$ from 1 to $\numnode_{ij}$ where $\numnode_{ij}$ denotes the
total number of nodes in $\SquaCent (\FirNode)$. We refer to the
label of node $\FirNode$ as $\Enum _ {\SquaCent (\FirNode) }
(\FirNode)$ where $\Enum_{\SquaCent(\FirNode)}(.)$ is the
enumeration operator for the square $\SquaCent (\FirNode)$. Also let
$\asn=\min_{i,j} \numnode_{ij}$ denote the minimum number of nodes
in one sub-square which by assumption is greater than
$a\log{\numnode}$ for some constant $a$. We split the problem into
three different cases. Figure \ref{FigPathRGG} illustrates these
there different cases.

\paragraph{Case 1} Nodes $\FirNode$ and $\SecNode$ do not belong to the the same
column or row. In this case, a two hop path $\Mpath _ {\FirNode
\SecNode} = (\FirNode \to \ThrNode \to \SecNode)$ is considered.
First, we pick $\SquaCent (\ThrNode)$, the vertex of the rectangle
constructed by  $\SquaCent (\FirNode)$ and $\SquaCent (\SecNode)$
with the same $x$-coordinate as $\SquaCent (\SecNode)$ and the same
$y$-coordinate as $\SquaCent (\FirNode)$. Now choose a node,
$\ThrNode$, inside $\SquaCent (\ThrNode)$ such that
\begin{equation}\label{eq:path_poincare}
\Enum _ {\SquaCent (\ThrNode)} (\ThrNode) = \Enum _ {\SquaCent
(\FirNode)} (\FirNode) + \Enum _ {\SquaCent (\SecNode)} (\SecNode)
\quad \text{mod}\ \asn.
\end{equation}
Since each square has at least $\asn$ nodes, such a choice can be
made. On the other hand, since nodes in each square is picked
uniformly at random in the averaging phase and there are at most
$b\log{n}$ nodes in each square (for some constant $b$) we have $
\bbW _ {\FirNode \ThrNode} , \bbW _ {\ThrNode \SecNode} \ge \frac
{1} {2m (b\log{n})^2} $, where the factor of 2 is due to the choice
of $\zeta$, the averaging direction. Substituting this inequality
into \eqref{EqnPoinPathWeight}, we obtain
\begin{align*}
| \Mpath _ {\FirNode\SecNode} | & = \frac{1} {\bbW _{\FirNode
\ThrNode} \StaDist (\FirNode)} + \frac {1} {\bbW _{\ThrNode
\SecNode} \StaDist (\ThrNode)} \; \leq \; 4 b^2 m \numnode \,
(\log{\numnode})^2.
\end{align*}
Furthermore, from equation~\eqref{eq:path_poincare}, we see that for
a fixed $s$ there are at most $\frac{b}{a}$ nodes in the square
$\SquaCent (\SecNode)$ that result in choosing $w$. Therefore, edge
$e:(\FirNode \to \ThrNode)$ is involved in at most
$\frac{b}{a}(m-1)$ such paths.

\paragraph{Case 2} Nodes $\FirNode$ and $\SecNode$ belong to the same row or column.
In this case, by setting $\Mpath _ {\FirNode \SecNode} = (\FirNode
\to \SecNode)$, we obtain
\begin{equation*}
| \Mpath _ {\FirNode \SecNode} | = \frac {1} {\bbW _ {\FirNode
\SecNode} \StaDist (\FirNode)} \le 2b^2mn(\log{n})^2.
\end{equation*}
Note that there is only one path containing $e$ of this type.

\paragraph{Case 3} Nodes $\FirNode$ and $\SecNode$ belong to the same square, meaning
$\SquaCent (\FirNode) = \SquaCent (\SecNode)$. In this case a node
$\ThrNode$ is chosen in a square adjacent to $\SquaCent (\FirNode)$
according to \eqref{eq:path_poincare} such that $\SquaCent
(\ThrNode)$ is to the right of $\SquaCent (\FirNode)$; unless
$\SquaCent (\FirNode)$ is in the last column, in which case
$\SquaCent (\ThrNode)$ is to the left of $\SquaCent (\FirNode)$. The
same argument as case 1 would give us a bound on $| \Mpath _
{\FirNode \SecNode} |$. As for the upper bound on the number of
paths: the edge $e : (\FirNode \to \ThrNode)$ is involved in at most
$\frac {b} {a}$ such paths.

Combining all the pieces, we obtain
\begin{equation*}
| \Mpath _ {\FirNode \SecNode} | \; \le \; 4b^2mn(\log{n})^2 \quad
\forall \; \FirNode,\SecNode \in \vertex,
\end{equation*}
and
\begin{equation*}
\max _ {e \in \NewEdges} \sum _ {\FirNode,\SecNode} \ind {\Mpath _
{\FirNode \SecNode} \ni e} \; \le \; m \: \frac{b}{a}+1.
\end{equation*}
Substituting these two inequalities into \eqref{EqnPoinCoef} yields
\begin{eqnarray*}
\PoinCoef
&\le& \big( m \: \frac{b}{a} \,  + \, 1 \big) \;  \frac {4b^2 \, mn \, (\log{n})^2} {n^2} \\
&\le& \frac {2mb} {a} \; \frac {4b^2 \, mn \, (\log{n})^2} {n^2}\\
&=&c_1\log{n}
\end{eqnarray*}
for some constant $c_1$. Therefore, from Poincare Theorem, we have
\begin{equation*}
\lam_2 (\SpecLapAve) = 1- \lam_{n-1} (\bbW) \ge \frac{1} {\PoinCoef}
\ge \frac{1}{c_1\log{n}}
\end{equation*}
which concludes the second part of Lemma \ref{LemEigenvalue}.

\subsection{Proof of part (a) of Theorem \ref{ThmMain}}

We now return to the proof of part (a) of Theorem \ref{ThmMain}.
Combining equations \eqref{EqnDecTheta} and \eqref{EqnUnrappedAlpha}
yields
\begin{align}\label{EqnThetaDecomposition}
\bth (\newt) \, = \, ( \thetacent \: - \: w(\newt) ) \: \bone \, +
\, \tbU \bbet (\newt),
\end{align}
where $w (\newt) = \frac{1} {\sqrt{\numnode}} \:
\sum_{l=0}^{\newt-1}\eps(l) \,
\inprod{\frac{\bone}{\sqrt{n}}}{\bv(l)}$. As previously established,
we know that $\Exp{w(\newt)} = 0$ and $\var{w(\newt)} \le \frac{\sig
\delta} {[\SpecGap] ^2}$ for all $\newt = 1, 2, \cdots$. Therefore,
invoking a result on convergence of series with bounded variance
(Theorem 8.3 from Chapter 1 of \cite{Durrett05}),  we have
\begin{align}\label{EqnConvAlpha}
w (\newt) \as w \quad \text{as $\newt \to \infty$},
\end{align}
for some random variable $w$. Since $w(\newt)$ is a sum of
independent Gaussian random variables (and hence Gaussian), it is
absolutely integrable \cite{Durrett05}. Therefore, we have $\Exp{w}
= \lim_{\newt \to \infty} \Exp{w(\newt)} = 0$ and also $\var{w} =
\lim_{\newt \to \infty} \var{w(\newt)} \le \frac{\sig \delta}
{[\SpecGap] ^2}$.

Now we move on to the next part of the proof, analyzing the sequence
$\{\bbet(\newt)\}_{\newt = 1} ^{\infty}$ using techniques from
stochastic approximation theory (e.g., see the
books~\cite{KusYin03,BenEtal90}). These techniques apply to
recursions that generate a state sequence $\{\bth(t)\}_{t=1}^\infty$
according to
\begin{align*}
\label{Eq:StocProc} \bth (t+1) \, & = \, \bth (t) \: - \: \eps(t) \:
H (\bth (t), v (t)) \quad t = 1, 2, \cdots,
\end{align*}
where $v(t)$ is the noise vector that models the randomness coming
into play in the algorithm.  The parameter $\eps(t)$ is a positive
step size, and the sequence $\{\eps(t)\}_{t=1}^\infty$ is required
to satisfy the conditions $\sum_{t = 1} ^ {\infty} \eps(t) = \infty$
and $\sum_{t = 1} ^ {\infty} \eps(t)^\alpha < \infty$ for some
$\alpha > 1$. The asymptotic behavior of these stochastic updates
can be analyzed in terms of the ordinary differential equation (ODE)
\begin{equation} \label{Eq:ODE}
\frac{d \gamma(\contt)}{d\contt} = - h(\gamma),
\end{equation}
where $h(\theta) \defn \Exs[H(\theta, v)]$. Under mild regularity
conditions, it is known that $\bth(t) \as \gamma^\ast$, where
$\gamma^\ast$ is the attractor of the ODE \eqref{Eq:ODE}.

Recalling the update equation \eqref{EqnBetaRecur}, our problem can
be cast within this framework.  In particular, the state sequence is
$\{ \bbet(\newt) \} _ {\newt=1} ^{\infty}$, the noise sequence is
formed by zero-mean i.i.d. random vectors, the decreasing sequence
is $\eps (\newt) = 1/(\frac{1}{\delta} + \newt)$, and finally
\mbox{$H(\bbet, v) = (\Ltil \bbet + \tbU^Tv)$} is a linear function
with $h(\bbet) = \Exs [\Ltil] \bbet$.  Note because we removed the
zero eigenvalue from the average Laplacian matrix, the matrix
$\Exs[\Ltil]$ has all positive eigenvalues, and so $\gamma^\ast = 0$
is the unique stable point of the linear differential equation
\mbox{$\frac{d \gamma(\contt)}{d\contt} = -\Exs[\Ltil] \gamma$}.
Therefore, an application of the ODE
method~\cite{KusYin03,BenEtal90} guarantees that
\begin{align}\label{EqnConvBeta}
\beta(\newt) \stackrel{a.s.}{\longrightarrow} 0 \quad \text{as
$\newt \to \infty$}.
\end{align}

Substituting the results \eqref{EqnConvAlpha} and
\eqref{EqnConvBeta} into equation \eqref{EqnThetaDecomposition}, we
obtain
\begin{align*}
\bth(\newt) \as (\thetacent - w) \bone \quad \text{as $\newt \to
\infty$}.
\end{align*}
In other words, nodes will almost surely reach a consensus;
moreover, the consensus value, $\ConsValue = \thetacent - w$, is
within $\frac {\sig \delta} {[\SpecGap] ^2}$ distance of the true
sample mean.


\section{Simulation Results}
\label{SecSimResult}

In order to demonstrate the effectiveness of the proposed algorithm,
we conducted a set of simulations.  More specifically, we apply the
proposed algorithm to four nearest-neighbor square grids of
different sizes.  We initially generate the data $\bth_i(0)$, $i =
1, 2, \cdots, \numnode$ as random $N(1,1)$ variables and fix them
throughout the simulation. So for each run of the algorithm the
initial data is fixed. In implementing the algorithm,  we adopt
$\sig = 1$ as the channel noise variance, and we set the tolerance
parameter $\delta = 0.1$, leading to the step size $\eps (\newt) =
\frac{1}{10 + \newt}$. We estimated the mean-squared error, defined
in equation \eqref{EqnMSEDefinition}, by taking the average over 50
sample paths.  As discussed in Section~\ref{SecMain}, every outer
phase update requires $\InnerIt = \bO{\sqrt{n}}$ time steps.

\begin{figure}
\begin{center}
\widgraph{.45\textwidth}{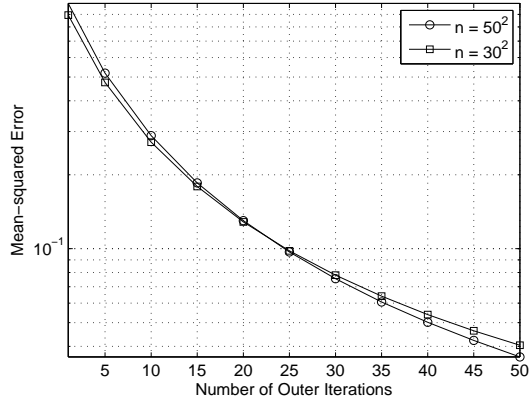}
\end{center}
\caption{Mean-squared error versus the number of outer loop
iterations for grids with $n \in \{ 30^2, 50^2 \}$ nodes. As
expected the MSE monotonically decreases, which supports  the
convergence claim.}\label{FigMSE}
\end{figure}

Figure~\ref{FigMSE} shows the mean-squared error versus the number
of outer loop iterations; the panel contains two different curves,
one for a graph with $\numnode = 30^2$ nodes, and the other for
$\numnode = 50^2$ nodes. As expected, the MSE monotonically
decreases as the number of iterations increases, showing convergence
of the algorithm. More importantly, the gap between the two plots is
negligible.  This phenomenon, which is predicted by our theory, is
explored further in our next set of experiments.

\begin{figure}
\begin{center}
\widgraph{.45\textwidth}{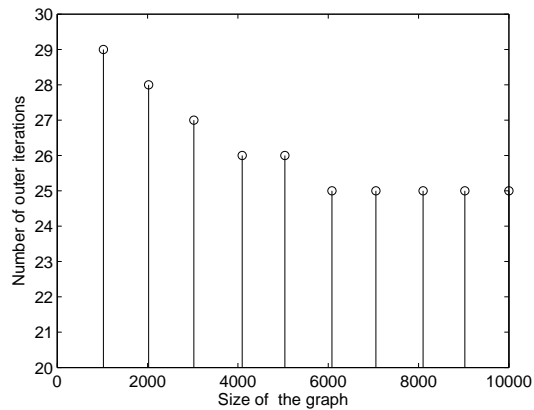}
\end{center}
\caption{Stopping time $\newt = \newt(n, \delta)$ vs. the graph size
$n$. For different graph sizes, we compute the first outer phase
time instance, $\newt (n, \delta)$, such that $\MSE (\bth (\newt
\InnerIt)) \le \sig \delta$. Here we have fixed the parameters to
$\sig = 1$, and $\delta = 0.1$. As you can see, over a range of
graphs of size varying from 1000 to 10000,  this stopping time is
roughly constant ($\approx 25$), which is consistent with the theory
(Theorem \ref{ThmMain}(b) and Lemma \ref{LemEigenvalue}).}
\label{Fig:GraphScaling}
\end{figure}

In order to study the network scaling of the grid more precisely,
for a given set of graph sizes, we compute the number of the
\emph{outer iterations} $\newt = \newt (n, \delta)$, such that $\MSE
(\bth (\newt \InnerIt)) \le \sig \delta$. Recall that this stopping
time is the focus of Theorem~\ref{ThmMain}(b).
Figure~\ref{Fig:GraphScaling} provides a box plot of this stopping
time $\newt$ versus the graph size $\numnode$.
Theorem~\ref{ThmMain}(b) predicts that this stopping time should be
inversely proportional to the spectral gap of the Laplacian matrix
$\SpecLapAve$, which for the grid scales as $\Omega(1)$ (in
particular, see Lemma~\ref{LemEigenvalue}).  As shown in
Figure~\ref{Fig:GraphScaling}, over a range of graphs of size
varying from $\numnode = 1000$ to $\numnode = 10000$, the stopping
time is roughly constant ($\newt \approx 25$), which is consistent
with the theory.

\section{Discussion} \label{SecConclusion}

In this paper, we proposed and analyzed a two-phase graph-respecting
algorithm for computing averages in a network, where communication is
modeled as an additive white Gaussian noise channel.  We showed that
it achieves consensus, and we characterized the rate of convergence as
a function of the graph topology and graph size.  For our algorithm,
this network scaling is within logarithmic factors of the graph
diameter, showing that it is near-optimal, since the graph diameter
provides a lower bound for any algorithm.

There are various issues left open in this work.  First, while the
AWGN model is more realistic than noiseless communication, many
channels in wireless networks may be more complicated, for instance
involving fading, interference and other types of memory.  In
principle, our algorithm could be applied to such channels and
networks, but its behavior and associated convergence rates remain to
be analyzed.  In a separate direction, it is also worth noting that
gossip-type algorithms can be used to solve more complicated types of
problems, such as distributed optimization problems
(e.g.,~\cite{NedicOz09,RamNeVe09,DucAgaWai10}).  Studying the issue of
near-optimal network scaling for such problems is also of interest.


\subsection*{Acknowledgements}
NN and MJW were partially supported by NSF grant CCF-0545862 from the
National Science Foundation, and AFOSR-09NL184 grant from the Air
Force Office of Scientific Research.


\bibliographystyle{plain}
\bibliography{Nima_STSP}

\begin{thebibliography}{10}

\bibitem{AyasoEtal}
O.~Ayaso, D.~Shah, and M.~Dahleh.
\newblock Information theoretic bounds for distributed computation over
  networks of point-to-point channels.
\newblock In {\em International Symposium on Information Theory}, 2008.

\bibitem{AysalEtal08}
T.~C. Aysal, M.~J. Coates, and M.~G. Rabbat.
\newblock Distributed average consensus with dithered quantization.
\newblock {\em IEEE Transactions on Signal Processing}, 56:4905--4918, 2008.

\bibitem{AyselEtal09}
T.~C. Aysal, M.~E. Yildiz, A.~D. Sarwate, and A.~Scaglione.
\newblock Broadcast gossip algorithms for consensus.
\newblock {\em IEEE Transactions on Signal Processing}, 57:2748--2761, 2009.

\bibitem{BenezitEtal10}
F.~Benezit, V.~Blondel, P.~Thiran, J.~Tsitsiklis, and M.~Vetterli.
\newblock Weighted gossip: Distributed averaging using non-doubly stochastic
  matrices.
\newblock In {\em Proc. IEEE International Symposium on Information Theory},
  2010.

\bibitem{BenEtal08_con}
F.~Benezit, A.~G. Dimakis, P.~Thiran, and M.~Vetterli.
\newblock Gossip along the way: order-optimal consensus through randomized path
  averaging.
\newblock In {\em Forty-Fifth Annual Allerton Conference on Communication,
  Control, and Computing}, Sep 2007.

\bibitem{BenEtal90}
A.~Benveniste, M.~Metivier, and P.~Priouret.
\newblock {\em Stochastic approximations and adaptive algorithms}.
\newblock Springer-Verlag, New York, 1990.

\bibitem{BoydEtal06}
S.~Boyd, A.~Ghosh, B.~Prabhakar, and D.~Shah.
\newblock Randomized gossip algorithms.
\newblock {\em IEEE Transactions on Information Theory}, 52:2508--2530, 2006.

\bibitem{CatSayed10}
F.~Cattivelli and A.~H. Sayed.
\newblock Diffusion {LMS} strategies for distributed estimation.
\newblock {\em IEEE Transactions on Signal Processing}, 58(3):1035--1048, March
  2010.

\bibitem{Chung97}
F.~R.~K. Chung.
\newblock {\em Spectral Graph Theory}.
\newblock American Mathematical Society, 1997.

\bibitem{DeGroot74}
M.~H. DeGroot.
\newblock Reaching a consensus.
\newblock {\em J. Amer. Stat. Assoc.}, 69:118--121, 1974.

\bibitem{Diaconis91}
P.~Diaconis and D.~Stroock.
\newblock Geometric bounds for eigenvalues of {M}arkov chains.
\newblock {\em Ann. Applied Probability}, 1:36--61, 1991.

\bibitem{DimSarWai08}
A.~G. Dimakis, A.~Sarwate, and M.~J. Wainwright.
\newblock Geographic gossip: {E}fficient averaging for sensor networks.
\newblock {\em IEEE Trans. Signal Processing}, 53:1205--1216, March 2008.

\bibitem{DucAgaWai10}
J.~Duchi, A.~Agawarl, and M.~J. Wainwright.
\newblock Dual averaging for distributed optimization: {C}onvergence analysis
  and network scaling.
\newblock Technical Report arXiv:1005.2012, UC Berkeley, May 2010.

\bibitem{Durrett05}
Rick Durrett.
\newblock {\em Probability: Theory and Examples}.
\newblock Thomson Learning, 2005.

\bibitem{FagZam07}
F.~Fagnani and S.~Zampieri.
\newblock Average consensus with packet drop communication.
\newblock {\em {SIAM} J. on Control and Optimization}, 2007.
\newblock To appear.

\bibitem{Grimmett}
G.R. Grimmett and D.R. Stirzaker.
\newblock {\em Probability and Random Processes}.
\newblock Oxford Science Publications, Clarendon Press, Oxford, 1992.

\bibitem{GupKum00}
P.~Gupta and P.~Kumar.
\newblock The capacity of wireless networks.
\newblock {\em IEEE Trans. on Inf. Theory}, 46(2):388--404, Mar 2000.

\bibitem{Hatano05}
Y.~Hatano, A.~K. Das, and M.~Mesbahi.
\newblock Agreement in presence of noise: pseudogradients on random geometric
  networks.
\newblock In {\em Proceedings of the 44th IEEE Conference on Decision and
  Control}, December 2005.

\bibitem{KarMoura09}
S.~Kar and J.~M.~F. Moura.
\newblock Distributed consensus algorithm in sensor networks with imperfect
  communication: link failures and channel noise.
\newblock {\em IEEE Transactions on Signal Processing}, 57(5):355--369, Jan
  2009.

\bibitem{KemEtal03}
D.~Kempe, A.~Dobra, and J.~Gehrke.
\newblock Gossip-based computation of aggregate information.
\newblock In {\em Proc. IEEE Conf. Foundation of Computer Science (FOCS)},
  2003.

\bibitem{KusYin03}
H.~J. Kushner and G.~G. Yin.
\newblock {\em Stochastic Approximation and Recursive Algorithms and
  Applications}.
\newblock Springer-Verlag, New York, 2003.

\bibitem{LopSayaed07}
C.~G. Lopes and A.~H. Sayed.
\newblock Incremental adaptive strategies over distributed networks.
\newblock {\em IEEE Transactions on Signal Processing}, 55(8):4064--4077,
  August 2007.

\bibitem{LopSayed08}
C.~G. Lopes and A.~H. Sayed.
\newblock Diffusion least-mean squares over adaptive networks: Formulation and
  performance analysis.
\newblock {\em IEEE Transactions on Signal Processing}, 56(7):3122--3136, July
  2008.

\bibitem{NazEtal09}
B.~Nazer, A.~G. Dimakis, and M.~Gastpar.
\newblock Neighborhood gossip: Concurrent averaging through local interference.
\newblock In {\em Proc. IEEE ICASSP}, 2009.

\bibitem{NedicOz09}
A.~Nedic and A.~Ozdaglar.
\newblock Distributed subgradient methods for multi-agent optimization.
\newblock {\em IEEE Transactions on Automatic Control}, 54:48--61, 2009.

\bibitem{Pen03}
M.~Penrose.
\newblock {\em Oxford studies in probability, Random Geometric Graphs.}
\newblock Oxford Univ. Press, Oxford U.K., 2003.

\bibitem{RajWai08}
R.~Rajagopal and M.~J. Wainwright.
\newblock Network-based consensus averaging with general noisy channels.
\newblock {\em IEEE Transactions on Signal Processing}, Jan 2011.

\bibitem{RamNeVe09}
S.~Sundhar Ram, A.~Nedic, and V.~V. Veeravalli.
\newblock Distributed subgradient projection algorithm for convex optimization.
\newblock In {\em IEEE International Conference on Acoustics, Speech, and
  Signal Processing}, pages 3653--3656, 2009.

\bibitem{HanGamal09}
H.~I. Su and A.~El Gamal.
\newblock Distributed lossy averaging.
\newblock In {\em Proc. IEEE International Symposium on Information Theory},
  2009.

\bibitem{Tsitsiklis84}
J.~Tsitsiklis.
\newblock {\em Problems in decentralized decision-making and computation}.
\newblock PhD thesis, Department of EECS, MIT, 1984.

\end{thebibliography}


\begin{biography}[{\includegraphics[width=1in,height=1.25in,clip,keepaspectratio]{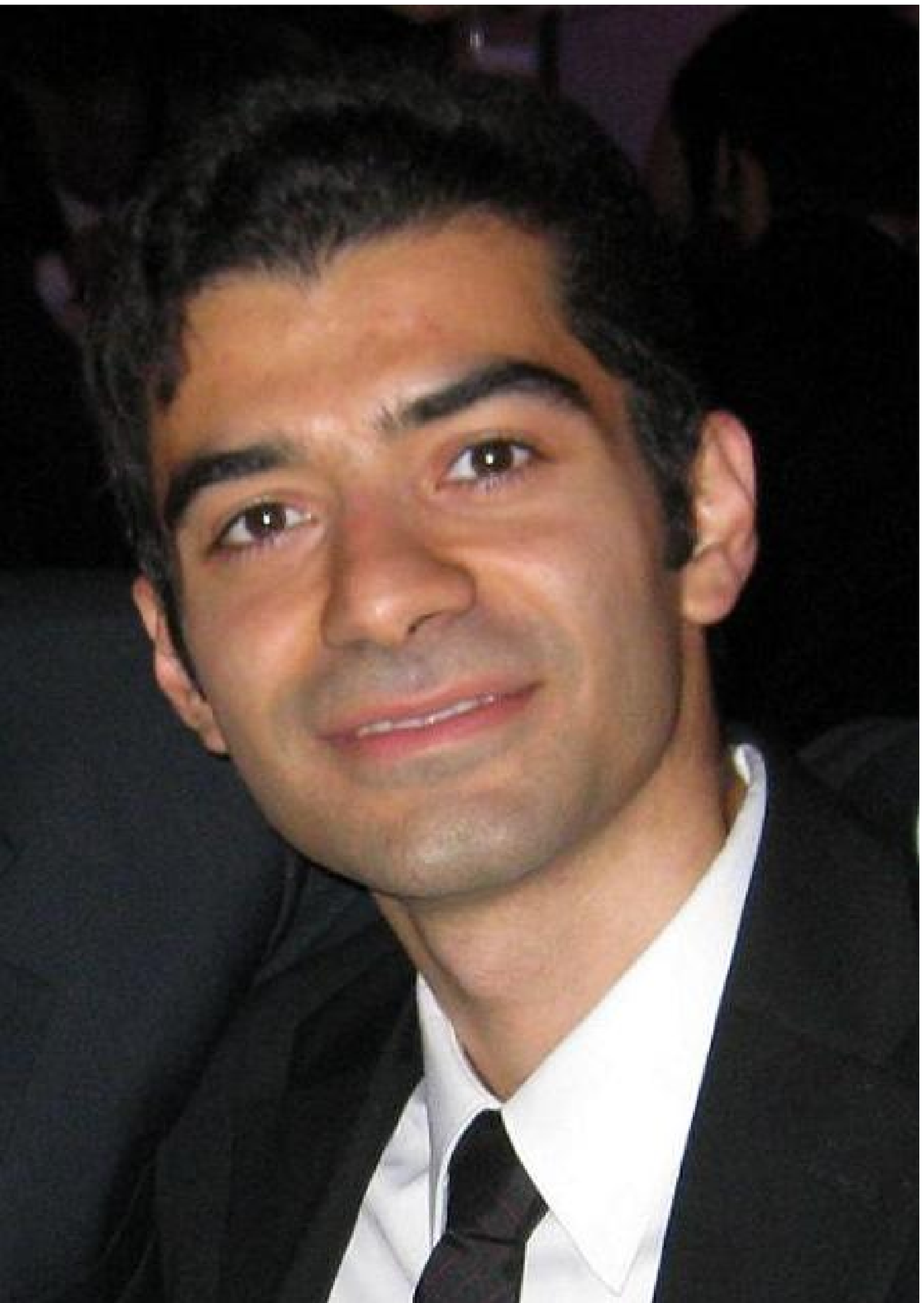}}]{Nima Noorshams}
received his B.Sc. from Sharif University of Technology, Tehran,
Iran, in 2007. He is currently pursuing his M.Sc. degree in the
department of Statistics and his Ph.D. degree in the department of
Electrical Engineering \& Computer Science at University of
California, Berkeley. His current research interests include
stochastic approximation methods, graphical models, statistical
signal processing, and modern coding theory.
\end{biography}

\begin{biography}[{\includegraphics[width=1in,height=1.25in,clip,keepaspectratio]{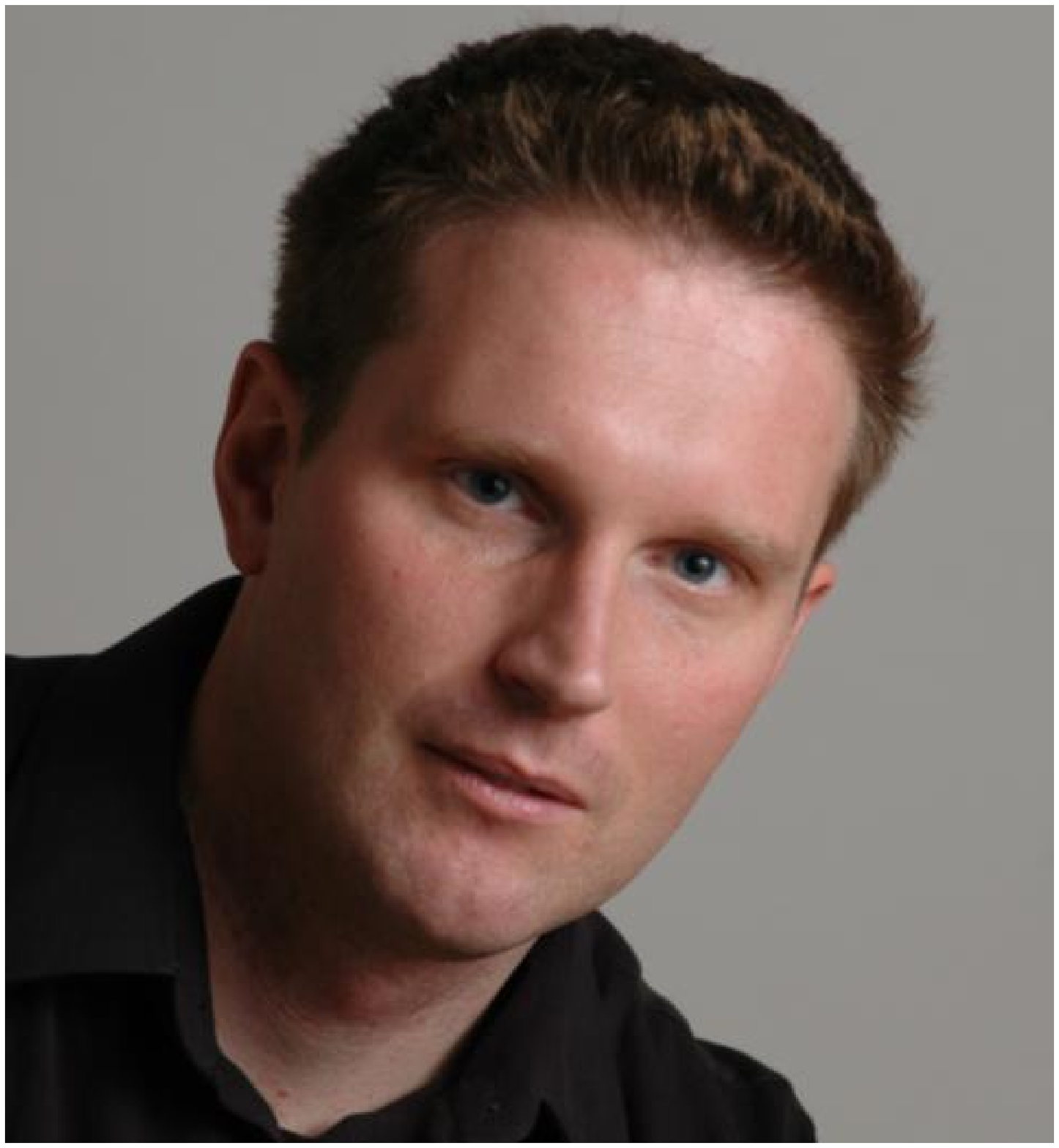}}]{Martin Wainwright}
 is currently an associate professor at University
 of California at Berkeley, with a joint appointment between the
 Department of Statistics and the Department of Electrical Engineering
 and Computer Sciences.  He received a Bachelor's degree in
 Mathematics from University of Waterloo, Canada, and Ph.D. degree in
 Electrical Engineering and Computer Science (EECS) from Massachusetts
 Institute of Technology (MIT).  His research interests include coding
 and information theory, machine learning, mathematical statistics,
 and statistical signal processing.  He has been awarded an Alfred
 P. Sloan Foundation Fellowship, an NSF CAREER Award, the George
 M. Sprowls Prize for his dissertation research (EECS department,
 MIT), a Natural Sciences and Engineering Research Council of Canada
 1967 Fellowship, an IEEE Signal Processing Society Best Paper Award
 in 2008, and several outstanding conference paper awards.
\end{biography}


\end{document}